\documentstyle[aas2pp4,psfig]{article}

\def\wisk#1{\ifmmode{#1}\else{$#1$}\fi}
\def\arcpt{\wisk{''\mkern-7.0mu .\mkern1.4mu}}

\newcommand{\ha}{H\,$\alpha$}
\newcommand{\hb}{H\,$\beta$}

\newcommand{\oii}{[O~{\sc ii}]}

\newcommand{\civ}{C~{\sc iv}}

\newcommand{\lya}{Ly\,$\alpha$}

\newcommand{\mgii}{Mg~{\sc ii}}
\newcommand{\nv}{N~{\sc v}}

\newcommand{\ewa}{$W_{\rm abs}$}

\newcommand{\aopt}{$\alpha_{\rm opt}$}
\newcommand{\arad}{$\alpha_{\rm rad}$}
\newcommand{\gl}{$\lambda$}

\newcommand{\zaze}{$z_{\rm a} \approx z_{\rm e}$}

\newcommand{\kms}{km\,s$^{-1}$}

\newcommand{\degree}{$^{\circ}$}

\slugcomment{To appear in Ap J.}

\lefthead{Baker et al.} 
\righthead{Associated C\,{\sc iv} Absorption in Radio-loud Quasars }

\begin{document}

\title{Associated absorption in radio quasars. I. 
\civ\ absorption and the growth of radio sources
}

\author{Joanne C. Baker\altaffilmark{1,2,3},
Richard W. Hunstead\altaffilmark{4},
Ramana M. Athreya\altaffilmark{5,6},
Peter D. Barthel\altaffilmark{7},
Eric de Silva\altaffilmark{3}, \\
Matthew D. Lehnert\altaffilmark{8},
Richard~D.E.~Saunders\altaffilmark{3} 
}

\authoremail{jcb@astro.ox.ac.uk}

\altaffiltext{1}{Hubble Fellow. }
\altaffiltext{2}{Astronomy Department, 601 Campbell Hall, 
	University of California, Berkeley, CA 94720, USA.}
\altaffiltext{3}{Astrophysics, Cavendish Laboratory, Madingley Road, 
        Cambridge, CB3 0HE, UK.}
\altaffiltext{4}{School of Physics, 
       University of Sydney, NSW 2006, Australia. }
\altaffiltext{5}{Institut d'Astrophysique de Paris, 98bis Boulevard Arago, 
	75014 Paris, France.}
\altaffiltext{6}{European Southern Observatory, Casilla 19001, 
	Santiago 19, Chile.}
\altaffiltext{7}{Kapteyn Institute, Postbus 800, 9700 AV Groningen, The Netherlands}
\altaffiltext{8}{Max-Planck-Institut f\"ur Extraterrestrische Physik,
Postfach 1312, 85741 Garching, Germany.	}


\begin{abstract}

We assess the global properties of associated 
\civ\gl\gl1548,1550 absorption lines measured in the 
spectra of radio-loud quasars drawn from a near-complete, 
low-frequency selected sample. The observations
span restframe \civ\ in two redshift ranges --- $0.7<z<1.0$ and 
$1.5<z<3.0$ --- which were targetted in the UV with HST/STIS and
in the optical with ground-based telescopes, respectively. 
First, we corroborate trends for \civ\ associated absorption 
to be found preferentially in steep-spectrum and 
lobe-dominated quasars, implying the absorbing material
tends to lie away from the radio-jet axis.
Furthermore, we find a clear anticorrelation between 
\civ\ absorption strength and the projected linear size of 
steep-spectrum quasars, indicative of an 
evolutionary sequence. 
We also find that heavily-absorbed quasars are systematically redder,
implying dust is mixed in with the \civ-absorbing gas. 
No redshift dependence was found in any of the trends considered.
These new results show that radio sources are triggered
in galaxies which are exceptionally rich in gas and dust, 
which then dissipates on a timescale comparable with, 
but less than, that of the radio source. 
This observational sequence, together with
the lack of redshift-dependence, points to a direct
causal link between the event which triggered the radio source and
the build-up of absorbing gas and dust, whose make-up 
is tantalisingly similar to the products of a post-merger starburst.
Thus, these new results provide direct evidence for the 
clearing of absorbing material around quasars with time,
as well as the probable association of starburst activity
with the onset of the radio activity in AGN.

\end{abstract}



\section{Introduction}
\label{sec:intro}

Quasar absorption lines provide a unique and powerful 
tool for probing tenuous gas clouds throughout the universe. 
The physical regimes probed in absorption differ from those 
which dominate the emission of both galaxies and Active Galactic 
Nuclei (AGN), and hence absorption provides a complementary 
view of the material composition of galaxies and AGN. 
Consequently, absorption lines occurring very close to the 
quasar emission-line redshift --- 
associated absorbers --- are valuable probes of 
local quasar environments, potentially intercepting gas from a range of
locations along the sightline. For instance, absorption can 
arise in gas flows near the quasar central engine, 
the interstellar medium (ISM) of the quasar host galaxy, 
or the ISM of neighbouring galaxies which may themselves 
be in clusters or groups. 

Associated absorption systems (usually defined by an absorption
redshift, $z_a$, lying within 5000\,\kms\ of the 
emission-line redshift, $z_e$) seem to be 
more numerous than can be accounted for
by randomly distributed galaxies along the 
quasar sightline (Weymann et al. 1979;  Foltz et al. 1986, 1988; 
Richards et al. 1999, 2001 and references therein; 
but see e.g. Young, Sargent \& Boksenberg 1982). The 
absorption systems occur most frequently in 
radio-loud quasars, especially those with steep radio spectra 
(Anderson et al. 1987; Foltz et al. 1988; Richards et al. 1999, 2001), 
favouring a direct link with AGN activity and also, perhaps, the processes 
of radio-jet formation. 

Many \zaze\ absorbers also differ subtly in ionisation or 
velocity  profiles from those seen along sightlines traversing 
normal galaxy halos, suggesting \zaze\ absorbing clouds 
experience directly the hard radiation field and kinematics of the AGN
(Petitjean, Rauch \& Carswell 1994; Hamann \& Ferland 1999; 
Hamann et al. 2001). 
In the most extreme cases, very broad absorption-lines (BALs; 
FWHMs of thousands of \kms) are seen
(see summaries by Turnshek et al. 1988; Turnshek 1995; Weymann 1995).
These broad troughs are probably due to outflowing,
dense winds close to the nucleus. Interestingly, BAL systems
occur almost exclusively in radio-quiet quasars (but see 
Becker et al. 1997, 2000; Gregg et al. 2000). 
This paper, however, will focus exclusively on the
narrow associated absorption-line systems (FWHMs tens to hundreds of \kms).

Among radio-loud quasars, the trend for \zaze\ absorption to 
occur preferentially in steep-spectrum  and
lobe-dominated quasars (Barthel, Tytler \& Vestergaard 1997) 
can be interpreted within the basic framework of the
`unified schemes' for AGN as an orientation effect.
Steep-spectrum radio sources are thought to be viewed at large
angles, $\theta$, to the radio-jet axis, and the presence
of absorbing material is consistent with the basic geometry of
an obscuring torus which blocks the nuclear light from the
observer beyond a viewing angle of about $\theta \sim 45$\degree 
(see Barthel 1989; Antonucci 1993; Urry \& Padovani 1995). 
In practice, considerable amounts of obscuring material must also 
lie within the torus opening angle, to cause the optical
reddening seen even in steep-spectrum quasars (Baker \& Hunstead 
1995; Baker 1997).

All properties of the AGN population cannot, however, be  
explained in terms of viewing angle differences alone. For example, 
20--30\% of {\it compact\/} radio sources (angular sizes $<2''$)
have steep rather than flat spectra, implying their radio jets 
are not beamed strongly towards us. 
Therefore, the projected sizes of compact, steep-spectrum (CSS) 
sources are not small simply because their jets
are oriented close to our line of sight, but rather they
are intrinsically small (see  Fanti et al. 1990).
In fact, as a class, CSS sources (see review by O'Dea 1998) do obey 
the orientation-unification picture (e.g. Saikia et al. 2001),
but they comprise intrinsically small 
versions of the larger radio-source population, their 
luminous radio emission emanating from regions within
the envelope of the host galaxy. 
The majority of CSS sources are thought to be young, and
presumably will grow into large-scale sources (Readhead et al. 1996; 
Owsianik \& Conway 1998; Taylor et al. 2000; Saikia et al. 2001;
de Silva et al., 2001, in preparation). Therefore source evolution
must also be folded into AGN studies. However, apart from the
evolution of the radio components, which has been extensively 
modelled, little is known about evolution of individual 
quasars at other wavelengths. 

To build up a picture of how the absorbing material is 
distributed around quasars, or its rate of occurrence
among different types of quasar, necessitates sampling
many sightlines over a wide range of viewing angles to 
the nucleus. Therefore, it is crucial to use complete 
and unbiased samples for such an investigation. 
Unfortunately, most previous work was based on 
heterogeneous samples. 
In this paper, we investigate associated \civ\ absorption 
in a complete sample of low-frequency selected quasars.
Using new optical spectra at moderate resolution 
(1--3\AA), we focus initially on the \civ\gl\gl1548,1550 
absorption doublet for a basic assessment of absorption 
characteristics. Spectra of the entire \lya\ to \civ\ 
restframe region and further analysis 
will be presented in a second paper (Paper II). 
To test for any redshift dependence of the absorbers
(e.g. Ganguly et al. 2001), quasars were observed in 
two redshift ranges ---
(i) $1.5<z<3.0$, where \lya\ and \civ\ are observable from the ground,
and (ii) $0.7<z<1.0$, where \lya\ and \civ\ have been observed with
STIS on HST. CSS quasars are included in this study --- we 
define them as having linear sizes less than $D=25$ kpc 
(measured between the outer radio hotspots)
and a steep spectrum ($\alpha>0.5$; 
where $S_{\nu} \propto \nu^{-\alpha}$)
as observed between 408\,MHz and 4.86\,GHz (see Kapahi et al. 1998). 
We also define the orientation-dependent ratio, $R$, to be 
the ratio of core to lobe luminosity
measured at a rest-frame frequency of 10\,GHz (see Kapahi et al. 1998).
For consistency with earlier papers we use cosmological 
parameters $H_0=50$\,\kms Mpc$^{-1}$, $q_0 = 0.5$ and 
$\Lambda = 0$ throughout. We note that, over the redshift
range under consideration, the resulting linear sizes are
consistent to within 10\% of the values using current best
estimates for cosmological parameters (namely 
$\Omega_m=0.3$, $\Omega_{\Lambda}=0.7$, $H_0=70$\,\kms 
Mpc$^{-1}$).

\section{Observations}
\label{sec:obs}

\subsection{The Parent Sample}

The quasars are all drawn from the well-defined 
Molonglo Quasar Sample (MQS; Kapahi et al. 1998). 
Briefly, the MQS comprises all quasars brighter than
$S_{408}=0.95$\,Jy
in the $-30^{\circ}< \delta < -20^{\circ}$ Declination
strip (excluding low Galactic latitudes, $|b|<20^{\circ}$) 
of the 408-MHz 
Molonglo Reference Catalogue (MRC; Large et al. 1981). 
Optical identifications for all 557
MRC radio sources in the strip above this flux-density limit 
have been obtained, and 111 quasars (including 6 BL-Lacs)
have been identified following spectroscopy. Redshifts of the
quasars span the range $0.1<z<3.0$. 
Due to its careful selection 
the MQS is estimated to be more than 97\% complete, with no 
optical magnitude limit, and is also relatively unbiased in terms of 
radio-jet orientation, as isotropic components dominate the 
radio emission at 408\,MHz.
Full details of the sample selection, and radio images, are given by 
Kapahi et al. (1998), and the initial low-resolution
optical spectroscopy is presented by Baker et al. (1999), 
to which the reader is referred. 
To study associated absorption systems in the quasars,
additional spectroscopy of the restframe \lya\ to \civ\ region has been
sought for two redshift-limited sub-samples, described below.
Tables \ref{hightab} and \ref{lowtab}
list these low- and high-redshift sub-samples, together
with basic radio and optical properties and observational details. 

\subsection{Ground-based spectroscopy of $z\sim 2$ quasars}

A program of moderate-resolution (1--2.4\AA\ FWHM)
spectroscopy of the \lya\ to \civ\ region was undertaken
for all 27 MQS quasars in the range $1.5<z<3.0$. To date,
observations have been completed for 20/27 of these quasars 
(see Table \ref{hightab}).
Two additional quasars have been observed at \lya\ 
only, namely MRC\,B0237--233 (a Gigahertz-Peaked Spectrum, GPS, source)
and MRC\,B0246--231 (CSS), 
and so are not considered in this paper. 
Of the remaining five quasars, four in the RA range
21h--23h, were missed due to weather-affected 
observing runs (two lobe-dominated, one core-dominated and 
one CSS quasar), and MRC\,B0522--215 (CSS) was not observed
on account of its faintness (and its uncertain quasar 
classification). 

The spectroscopy was
carried out mostly with the Anglo-Australian Telescope (AAT) 
at Siding Spring, Australia, and the ESO 3.6m telescope at La Silla, Chile,
in several runs between 1995 and 2000. In addition, four faint quasars 
were observed in 1999 with FORS1 on the VLT UT1 at Paranal, Chile. 
We note that these observations were challenging due to the faintness of
the targets ($19<b_{\rm J}<22$), 
requiring exposures of several hours each at the highest dispersion
with 1200 line/mm gratings, and at least an hour per target in good seeing
with 600 line/mm gratings. 

Observations at the AAT were carried out in four runs on UT dates
1995 October 19-20, 1996 March 21-22, 1997 February 9-10 and
1999  April 11-12. 
For the 1995 and 1996 runs we used the 
RGO spectrograph at f/8 with the 25-cm camera and 
Tek $1024\times 1024$ CCD and 
high-dispersion 1200V grating oriented blaze-to-camera. 
Observations were taken with a 1.5--2\arcsec\ slit (comparable with seeing)
oriented at parallactic angle. The resulting spectral 
resolution was 1.2\AA\ FWHM.  
At this dispersion it was necessary to target 
\civ\ and/or \lya\ (when visible) in separate exposures. 
Standard star and Cu-Ar comparison lamp spectra were taken at each 
grating setting. Wavelength calibration is accurate to $0.1$\AA\ rms.
Conditions were generally not photometric, although the 
flux densities are reliable to within 30\%.
All observations were taken at low airmasses where 
possible ($<1.2$); no extinction corrections were applied. 

In the 1997 and 1999 runs, we used the same setup with a 600V grating, 
giving a spectral resolution of 2.4\AA\ FWHM.
Typical seeing for both runs was 1\arcpt2--1\arcpt 3, and we used 
a slit 1\arcpt5 wide, oriented at parallactic angle.
Again, wavelength calibration was carried out to within 0.2\AA\ rms 
with the Cu-Ar lamp spectrum, and spectrophotometric standard stars 
were observed each night. Conditions were mostly clear for the 1999
run, but cirrus may have affected the 1997 data.
No extinction corrections were applied to observations made 
within $35^{\circ}$ of the zenith (i.e. at airmasses smaller than $1.2$).

On the ESO 3.6m telescope, EFOSC-2 was used on the nights of 
2000 February 1 and 2. The seeing was generally good ($\sim 1''$),
and a slit width of 1\arcpt 2 was used for the majority of
observations (reduced to $1''$ when the seeing was especially good). 
The slit was placed on the sky at parallactic angle. EFOSC grating
\#7 was used with a 2048-pixel square detector and 2-pixel binning, 
to give a mean dispersion of 1.0\AA\ per pixel over the wavelength 
range 3250--5235\AA. Helium-Argon calibration lamp spectra were observed 
for each object, giving a wavelength accuracy of better than 0.1\AA. 
A spectrophotometric standard star was observed twice nightly. 
Some objects were observed at large airmasses (1.2--2.0),
so extinction corrections were applied after the data were
reduced in these cases. 
The typical exposure time for each quasar was 1 hour.

Observations on the VLT UT1 telescope were made in service-observing mode 
on  1999 October 17, November 11-13, 15 and December 11-12. The FORS1
instrument was used with a 300B grism, giving a dispersion of 1.2\AA\ 
per pixel over the wavelength range 3000--7000\AA. The slit was $1''$ wide
and placed at parallactic angle, and the seeing was always 
better than $1.2''$. 
Spectrophotometric standard stars were observed each night. 
Two-dimensional wavelength solutions were applied to the raw spectral
data from archival calibration data. The linearised images were 
then cleaned of cosmic rays, bias-subtracted and flat-fielded 
before the spectra were extracted. 
Again the observations were made at low airmasses so extinction 
corrections were not applied. 

All the spectra were reduced using standard procedures in FIGARO,
and have been converted to a vacuum-heliocentric wavelength scale.

\subsection{HST observations}
\label{sec:hstobs}

For direct comparison with the $z\sim 2$ sample, 
UV spectroscopy was carried out with the STIS instrument on HST
for 19 MQS quasars with redshifts $0.7<z<1.0$.
More precisely, the HST targets comprise all MQS quasars in the redshift
range $z=0.780$--0.950, with the addition of B0123--226 and B1224--262 
(with similar redshifts of $z=0.717$ and 0.77, respectively). 
MRC\,B0123--226 was chosen to increase the number of 
core-dominated quasars (to three) in the HST target list (it has $R=2.10$,
albeit $\alpha = 0.58$, in Table \ref{lowtab}), 
and MRC\,B1224--262 was added to increase the number of CSS quasars
(to four), for statistical reasons. Given the already small
number of quasars targetted, both are included in the rest 
of the analysis (see also Section \ref{sec:discrad}). 
Two sources within the complete
redshift-limited sample were not targetted with HST --- BL-Lac 
object B2240--260, and quasar B0418--288 which was too faint 
($b_{\rm J}>21$) to observe with STIS. 
All targets were detected successfully with STIS, apart from 
B2156--245, which is both faint ($b_{\rm J}=20.2$) and red (\aopt $=2$), 
and is therefore excluded from this analysis.

The HST STIS observations were scheduled at various times 
between May 1999 and February 2001. The NUV-MAMA detector was 
used with the G230L grism centred at 2376\AA, 
giving a spectral resolution of 3.0\AA\ over the wavelength 
range 1570--3180\AA. The MAMA detectors have 
the double advantage of absence of read noise and lack of 
susceptibility to cosmic rays. 
A slit of length $52''$ and width 0\arcpt 2 was used for the observations. 
A total of 34 HST orbits was used, with one or two orbits per 
target depending on its brightness. Exposures in each orbit were split 
into two, of typical duration 1200s.
Standard wavelength and spectrophotometric calibrations
were used, and the data were reduced using the standard STIS pipeline.   
Table \ref{lowtab} details the observations for each target. 

\section{The spectra}
\label{sec:spectra}

The \civ\ restframe region is shown for both high- and low-redshift
quasars in Figures \ref{grdspectra} and \ref{hstspectra}, respectively. 
Where more than one ground-based  spectrum was taken for the former, 
only the spectrum with the highest signal-to-noise ratio is shown 
in Figure \ref{grdspectra}. 
Clearly, the \civ\ profiles shown in the Figures 
display a wide range of  absorption properties. 

Absorption systems were identified and measured in the following way
using `splot' in IRAF for both sets of spectra. 
Measurements are given in Tables \ref{datatab} and \ref{hstdatatab}. 
First, absorption systems were identified in each spectrum
lying within 5000\,\kms\ of the \civ\ emission-line redshift. 
The \civ\ emission-line redshift was measured by fitting a Lorentzian
profile to the line wings and using the fitted peak wavelength. 
Velocities of the narrow absorption lines are defined in the 
standard way, i.e. $v/c = (A^2 - 1) / (A^2 + 1 )$
where $A = (1 + z_e) / (1 + z_a)$ (Foltz et al. 1986),
although in practice we approximated this to 
$\Delta v / c \approx (z_e-z_a) / (1+z_e)$ (for $\Delta v\ll c$).
Identification of the \civ\ absorption lines was confirmed by identifying
other lines in the same system, usually 
including \lya\ (see Paper II for identifications of other species), 
and the ratio of wavelengths of 
the \civ\ doublet when it was resolved. In this way we selected the 
strongest \civ\ absorption system within 5000\,\kms\ for further 
measurement. Velocities were measured using the stronger 
1548\AA\ absorption line, 
or assuming an average wavelength of 1550\AA\ if the doublet was blended. 
Uncertainties in velocities are dominated by errors in fitting the
broad emission-line peaks, which can be uncertain by
several \AA. Errors are also greater for blended absorption lines.  
Velocity uncertainties are described more
fully in Sections \ref{sec:relvel} and \ref{sec:sysvel}. 
In four quasars with ground-based spectra, we also noted and measured
some weak absorption systems with larger relative velocities,
$\Delta v = 5000$--8000\,\kms, but these were ultimately 
excluded from the main 
analysis (see Section \ref{sec:relvel}).

Equivalent widths (\ewa) were measured for both lines of the 
\civ\gl\gl1548,1550 absorption doublet together, so that blended and
unblended lines could be compared. The measurements were made 
using a simple linear fit across the continuum adjacent 
to the absorption feature; this was deemed sufficient given the
absorption lines are narrow and superposed on the wings of broad emission
lines of different shapes. Given the range of emission profiles observed
(ranges of widths, `peakiness', and asymmetries), 
there is some subjectivity in continuum placement for the
equivalent width measurement, particularly when the absorption occurs
close to the line peak, in which case the absorption may be underestimated. 
The \ewa\ measurement errors quoted span
the range of reasonable fits to the data by hand,
taking continuum-fitting into account, and 
are typically $\sim 10$\% for the ground-based observations and 
$\sim 20$--30\% for the HST observations. The equivalent widths
are all quoted in the restframe. Upper limits ($\sim 3\sigma$) 
to \ewa\ were estimated by eye, corresponding roughly to the 
largest EW measurable from a noise feature within 5000\,\kms\ of 
the emission-line. Thus, they depend on the quality of the 
individual spectra, the continuum magnitude, and the strength 
and profile of the \civ\ emission lines, but are typically 
\ewa $<0.2$\AA\ for the groundbased and \ewa $<0.3$\AA\ for the
HST data (with three noisy HST spectra yielding \ewa $<0.8$\AA).

The statistical significance of correlations has been estimated throughout 
using the Kendall's Tau and Spearman's Rho rank correlation tests as coded in 
the ASURV program (Isobe \& Feigelson 1985, 1986), incorporating
upper limits in equivalent widths when relevant. Limits in $R$ and 
$D$ were not incorporated for the few objects affected due to 
software limitations, which restrict the application of multiple sets of
limits to the data. 

\subsection{Notes on individual HST spectra}
\label{sec:hstnotes}

The HST data have lower signal-to-noise ratios and resolution 
than the ground-based observations, so they are subject to somewhat greater 
uncertainties in line identification and measurement. 
Notes are given below for measurements of several quasars, 
as indicated in Table \ref{hstdatatab}.  

{\bf B0030--220:} the associated \civ\ absorption system 
is complex, and is almost certainly blended with Galactic 
\mgii\gl\gl2796,2803 at $z\approx 0$. 
However, the strength of the absorption system, together with the 
presence of associated \lya\ and \nv\ absorption, suggests that
associated \civ\ is primarily responsible. A large, conservative 
error bar has been assigned to the measured \ewa\ to try to
account for the uncertainty due to blended MgII, and the fact 
that the absorption feature lies close to the emission-line peak. 

{\bf B0135--247:} Galactic \mgii\ absorption  
is obvious in the blue wing of the \civ\ emission line.

{\bf B1202--262:} Galactic \mgii\ absorption is  visible 
in the red wing of the \civ\ emission line.

{\bf B1208--277:} Both the \lya\ and \civ\ emission lines 
are narrow, suggesting that the quasar classification of 
this object is uncertain, and instead it lies on the border-line 
between a radio galaxy and quasar. 
The optical spectrum published by Baker et al. (1999) is noisy, 
but it does not show any strong broad lines and \oii\ is 
weaker than expected in a classical radio galaxy spectrum, 
consistent with this interpretation. 
No absorption is detected against the narrow \civ\ emission 
line, but the likelihood of detection is much lower against 
such narrow lines and weak, noisy continuum. Therefore, given 
these uncertainties
we exclude B1208--277 from the rest of the analysis. 

{\bf B1349--265:} the HST spectrum is very noisy, but strong absorption is
clearly seen in \lya, \nv\ and \civ. There appear to be at least two sets of 
strong \civ\ doublets --- we have measured only the strongest one here. 

{\bf B2136--251:} the HST spectrum is noisy, but we note that the ratio of
\nv/\lya\ emission is unusually high, and the emission lines
are relatively narrow compared with the other quasars observed. 
Kapahi et al. (1998) classified this quasar as a GPS 
radio source, and it is the most compact
radio source in our sample. 

\newpage

\section{Results}

\subsection{Relative velocities}
\label{sec:relvel}

Figure \ref{vels}  shows the distributions of velocities 
relative to the \civ\ emission-line redshift for all the 
\zaze\ absorption systems identified in the groundbased and HST spectra, 
respectively. In both sub-samples the majority (50--70\%)
of strong absorption systems lies within 500\,\kms\ of 
the emission redshift. At slightly higher 
relative velocities, between $\sim 1000$--3000\,\kms\ 
of the emission redshift, 
we find both blue- and redshifted systems. There is a slight
tendency to find more blueshifted systems, but 
based on so few objects it is not statistically significant. 
The means of the  velocity distributions of the absorption systems 
(in Tables \ref{datatab} and \ref{hstdatatab})
are consistent with no offset from the \civ\ emission-line redshift 
($\langle \Delta v \rangle \approx  40$, 
80\,\kms\  for the high- and low-redshift datasets, respectively). 
The standard deviations in the distributions of measured velocities are 
both $\sim 1000$\,\kms\ ($\sigma \approx 1300$ and 800\,\kms, 
respectively; the difference is not significant). 

In the ground-based data, where the signal-to-noise ratios are higher, 
we detected four \civ\ absorption systems 
in the emission-line wings with still higher relative velocities,
$\Delta v = 5000$--10000\,\kms\ (see Table \ref{datatab}). These are exclusively 
blue-shifted systems, and all four are seen in core-dominated quasars,
as indicated in Figure \ref{rvel}. Possible explanations for
seeing such systems only in highly-beamed, core-dominated quasars
include (1) geometrical effects, such that we are intercepting 
faster outflowing gas near the radio-jet axis, or (2) these
absorbing clouds arise in foreground galaxies along the 
quasar sightline.
The latter interpretation is 
consistent with the high-velocity absorption lines being relatively weak
($0.5< W_{\rm abs} < 1.5$\AA) and the lack of any increase in 
relative velocity with radio core-dominance parameter, $R$, otherwise
in Figure \ref{rvel}. In this case the preferential detection of
these high-velocity systems in core-dominated quasars may simply be
a selection effect, whereby weak 
absorption systems will be easier to detect against the 
brighter continua of highly-beamed quasars, which in addition 
tend not to show any strong \zaze\ absorption (see Section \ref{sec:rdep}). 
These four quasars are among the most optically-luminous in our study (only two 
other quasars have comparable optical luminosities).
Similar high-velocity systems in core-dominated quasars have been
reported by Richards et al. (1999), and several other examples of 
quasar absorbers at extremely high relative velocities are known
(e.g. Hamann et al. 1997), but the relationship of this class 
to either low-velocity or intervening absorbers is unclear at present. 

Therefore, we will consider only the class of associated absorption systems 
with $\Delta v < 5000$\,\kms\ for the rest of this paper. 
We note that the inclusion of the four weak, high-velocity absorbers
would not, in fact, change any of the main conclusions. 

\subsubsection{Systematic uncertainties in velocity}
\label{sec:sysvel}

One uncertainty in the interpretation of relative 
velocities is that the \civ\ emission line may
not be representative of the systemic redshift of the quasar. 
It has been noted before that high-ionization 
lines in quasar spectra, including \civ,  may be blueshifted 
by up to $\sim 1000$\,\kms\
relative to lower ionization and narrow forbidden
lines, such as \mgii\ and \oii\gl3727 (Gaskell 1982; Tytler \& Fan 1992; 
McIntosh et al. 1999).
To check whether the velocities measured relative to the \civ\ 
emission-line redshift are biased with respect to the systemic 
redshift of the quasar, we plot in Figure \ref{velcomp} 
the relative
velocities of the absorption systems calculated relative to the 
\civ\ and \mgii\ emission lines, when both were available 
(Baker et al. 1999; de Silva et al., in preparation). 
Velocities relative to \oii\ are also included 
for several low-redshift quasars where this line is visible
in the optical spectrum. The agreement between the \civ\ and \mgii\ 
redshifts for the ground-based data in Figure  \ref{velcomp} 
is in general good (within 500\,\kms), except in two cases 
(B0328--272 and B2158--206) where \civ\ is blueshifted relative 
to \mgii\ emission by $\sim 1000$\,\kms. The apparent redshift difference
in B2158--206 may, however, be overestimated due to absorption in 
\mgii\ (Baker et al. 1999). The scatter for the $z\sim 1$
quasars is somewhat greater, as expected given the lower resolution
of the HST data (3\AA\ corresponds to about 300\kms\ resolution 
for \civ\ at 2800\AA, and the wavelength calibration for the 
MAMA detectors themselves is uncertain to within a pixel, or 
about 150\kms) and measurement uncertainties
for the emission-line peaks. In fact, the uncertainties in 
the relative velocities are
dominated by the uncertainty in fitting the broad emission-line
profile, which is exacerbated by the presence of the absorption itself
and also intrinsic profile asymmetries.
Thus, uncertainties in $\Delta v$ relative to the mean 
\civ\ emission redshift are $\pm 500$--600\,\kms\ for both
low- and high-redshift datasets, $\pm 700$--800\,\kms\ 
with respect to \mgii\ in the high-redshift subset and 
as high as $\pm 1000$\,\kms\ for the low-redshift quasars.
The higher values for \mgii\ are due to the lower-resolution of 
the spectra from which these wavelengths were measured.
These uncertainties are seen in the scatter in  Figure \ref{velcomp}.
Thus, apart from the two cases above, there is no strong evidence for a 
systematic velocity offset between the \civ,  \mgii\ and  
\oii\ emission lines.

Given not all objects have \mgii\ and/or \oii\ 
emission measurements available, and the effect of a small
zeropoint velocity shift is unimportant for the main conclusions 
of this study, we will continue to refer to velocities relative 
to the \civ\ emission-line redshift throughout the rest of the paper.

\subsection{Absorption as a function of radio spectral index}
\label{sec:alpha}

Figure \ref{bothewarad} shows the equivalent width of the 
\civ\ absorption lines as a function of 0.408--5\,GHz 
radio spectral index 
for both the high- and low-redshift datasets.  Strong
absorption (\ewa$>1$\AA) is detected {\it exclusively \/} 
in steep-spectrum ($\alpha > 0.5$) quasars in both samples. 
Statistically, a significant correlation is present, according to 
both Kendall's Tau and Spearman's Rho rank correlation tests, 
at the 95\% level
for the combined datasets. However, looking at the figure, 
the data seem to define a trend in the upper envelope of
the distribution of points, with considerable
scatter, rather than a direct linear relationship between the two 
quantities.

The MQS contains a relatively small number of flat-spectrum quasars 
due to its low-frequency selection. 
Nevertheless,  the lack of any strong absorption 
in the six quasars with spectra flatter than
0.5  is marginally significant, i.e. on the null hypothesis, 
we would otherwise expect about half of them (three) to 
exhibit strong absorption and none does. Although 
our study is limited by 
the small number of flat-spectrum quasars, 
the prevalence of strong absorbers in steep-spectrum quasars 
is in good agreement with earlier studies 
(Anderson et al. 1997; Foltz et al. 1988).

\subsection{Absorption as a function of radio-core dominance}
\label{sec:rdep}

The equivalent width of the \civ\ absorption 
lines is plotted as a function of radio-core-dominance parameter, $R$, in 
Figure \ref{rew}.  
CSS quasars are included in a separate panel in Figure \ref{rew} for 
comparison because $R$-values were not measurable on account of the
small sizes of these sources and the inherent difficulty in 
identifying core components in them.
Together, both the high- and low-redshift samples indicate
a trend in Figure \ref{rew}, such that absorption is 
greatest in lobe-dominated quasars of low $R$ and decreases 
steadily with $R$. The anticorrelation is significant at the
98\% level.  A similar trend was reported by Barthel et al. 
(1997) and interpreted in terms of the anisotropic distribution 
of the absorbers, lying predominantly away from the radio-jet axis.

We note the presence of two apparent outliers in Fig \ref{rew},  
B1355--215 and B0136--231 (labelled Q1 and Q2 on
Figure \ref{rew}, respectively). 
These both have somewhat uncertain $R$ values by 
about a factor of two, based on their radio images, and both 
have steep spectra (see Kapahi et al. 1998). 
B1355--215 is a small source (35\,kpc), only slightly larger than
the CSS definition used in this paper. That an occasional 
absorbing cloud intercepts the
line of sight to the quasar even when it is observed close to the 
jet axis is perhaps not unexpected, especially in smaller sources
(see below).

\subsection{Absorption as a function of radio-source size}
\label{sec:size}

The equivalent widths of \civ\ absorption  are plotted 
as a function of the projected linear sizes of the radio sources 
in Figure \ref{ewl} 
for both high- and low-redshift datasets. Core-dominated quasars
are now excluded as they are severely foreshortened. CSS quasars
are included --- linear sizes were measured from 1.7-GHz and
5-GHz MERLIN images with typical resolution 0\arcpt 1
(de Silva et al. 2001, in preparation).  Linear sizes
for larger sources are taken from 5-GHz VLA images at $1''$ resolution
(see Kapahi et al. 1998). Linear sizes are therefore accurate to
better than $\sim 10$\% in most cases. 

It is clear in Figure \ref{ewl}
that the strongest absorption occurs preferentially in the 
smallest sources. 
Notably, all but one of the CSS sources show \zaze\ absorption
stronger than $W_{\rm abs}=1$\AA. The exception is 
MRC\,B2136--251, a very compact GPS source ($D < 0.2$ kpc) with an 
unusual UV spectrum (see notes in Section \ref{sec:hstnotes}). 
The formal probability of an anticorrelation
being present is $>99$\% for the combined datasets 
(excluding the GPS source).

Figure \ref{ewl} demonstrates that, at least in this regard, 
CSS quasars do not appear to be a class separate from other quasars.
Instead, there is a continuous decline in absorption strength
as the sources increase in size. Thus, the  precise size limit
used to define CSS quasars, here 25\,kpc (chosen before this
absorption-line study was made; see Kapahi et al. 1998), 
is certainly quite arbitrary, although it corresponds approximately
to the typical scale of galaxies (tens of kpc).
Nevertheless, it does seem an appropriate limit in that it marks the 
size below which {\it all\/} MQS quasars (apart from GPS source 
MRC\,B2136--251) show significant optical absorption.
Although the cutoff size is irrelevant in plots such as
Figure \ref{ewl},  any comparison of average properties between classes,
such as CSS and larger sources,  would be
diminished in significance proportionately as larger cutoff 
sizes were used.

In Figure \ref{ewl}, both the high- and low-redshift datasets fall 
on the same relationship to within a factor of $\sim2$. 
We note that this need not be the case intrinsically, as
different cosmological models predict different
scaling relationships between observed angular size and projected linear
size, and merely the assumption of one set of cosmological
parameters rather than another would allow the high- and 
low-redshift datasets to be moved laterally with respect to 
each another on Figure \ref{ewl}. Consequently, we can 
say merely that high- and low- redshift MQS quasars follow 
the same relationship between absorption strength and linear size
for models with angular-size redshift relations similar to the 
cosmology assumed in this paper.
We also note that both high- and low-redshift datasets are 
indistinguishable in all the other correlations we have presented,
which are not cosmology-dependent.

In Figure \ref{lvel}, the velocities of the \civ\ absorption
lines relative to the broad \civ\ emission redshift are 
plotted as a function of radio linear size. There is a weak 
anticorrelation, such that \civ\ absorption tends to be 
blueshifted more often in small sources and may be occasionally 
redshifted in larger sources, but this is highly speculative 
given the large uncertainties involved (of order $\pm500$--600 
\kms\ due to measurement) 
and the statistically small number of quasars with measured 
absorption. Therefore we merely mention it here, until 
follow-up observations are made.
For comparison, no trend was found between
relative velocity and $R$ (Figure \ref{rvel}).

\subsection{Reddening by absorbing clouds?}
\label{sec:dust}

For quasars where  \civ\ absorption was detected, the equivalent
width of the absorption is plotted in Figure \ref{ewaopt} against 
the power-law slope of the optical continuum, $\alpha_{\rm opt}$ (as 
measured between 3500 and 10000\AA). There is a remarkably strong 
correlation (with a probability $>99$\%) 
between absorption-line strength and spectral 
slope --- heavily absorbed quasars are systematically redder. 
The unequivocal correlation argues strongly for a direct
relationship between the two, in the sense that \civ\ absorption
implies a redder continuum. 

Baker (1997) presented evidence that the range in optical spectral slope 
observed in the MQS is due in part to reddening by a dust 
screen lying outside the broad emission-line region. 
In this earlier study, the most direct evidence for dust reddening 
(as opposed to intrinsic spectral steepening) was the tight correlation 
between \aopt\ and the broad-line \ha/\hb\ Balmer Decrement,  
at least in low-redshift ($z<0.5$) quasars where it was 
measurable in the optical.
By extension, the simplest explanation of the 
correlation of $W_{\rm abs}$ with \aopt\ in Figure \ref{ewaopt} is that
the absorbing gas clouds also contain dust, and they lie well
outside the nuclear continuum source. In addition, the fact that
some quasars without strong \civ\ absorption have \aopt$>1$ 
indicates that the dust may have a greater covering fraction than 
the \civ-absorbing clouds, although absorbed quasars are 
systematically redder than unabsorbed quasars. 
Alternatively, if dust is not responsible, then \civ\ 
absorption strength correlates with an intrinsically 
softer continuum shape. 

No significant correlations were found
between $W_{\rm abs}$ and optical luminosity or magnitude. 
However, most of the quasars are concentrated in a relatively small 
range in B-band optical luminosity, $23.5 < \log L_{\nu} < 24.5$,
within the whole range   $22.3 < \log L_{\nu} < 25.2$ 
(where $L_{\nu}$ is expressed in WHz$^{-1}$).

\section{Discussion}

\subsection{Orientation and evolution}
\label{sec:discrad}

Evidence was presented in Sections \ref{sec:alpha}
and \ref{sec:size} that the strongest \civ\ absorbers 
are found more often in lobe-dominated, steep-spectrum quasars, 
but particularly in {\it small\/} radio sources. In fact, as
Table \ref{restab} shows, strong associated absorption
(\ewa$>1$\AA) was detected in a total of 11/12 ($92\pm8$\%) 
CSS (and GPS) quasars, compared with 8/19 ($42\pm11$\%) large 
steep-spectrum sources ($D>25$ kpc) and 0/6 flat-spectrum
sources (using binomial uncertainties). 
The median equivalent width
of \civ\ absorption is also correspondingly stronger in CSS quasars
than lobe-dominated quasars, and weakest in flat-spectrum sources. 

Although the redshift-limited samples are slightly incomplete,
the numbers of missing quasars (totalling 5 CSS, 2 lobe-dominated, 
1 core-dominated, and 1 GPS quasar) are too small to overturn the above
conclusions (and the inclusion of two quasars in the HST
sample which are technically outside the redshift-limited subset
redresses the balance for lobe-dominated quasars; see 
Section \ref{sec:hstobs}). We note that
the CSS quasars missed include some of the reddest, and 
faintest, quasars in the sample, and these very properties 
are consistent with the thrust of our results. 
Therefore, we predict that the majority
of these unobserved CSS quasars will show strong 
\civ\ absorption.

We now discuss whether the above results may be explained by 
one or a combination of the following hypotheses:
(1) the distribution of absorbing clouds is orientation-dependent;
(2) absorption strength correlates with the density of 
the gaseous (or cluster) environment surrounding the quasar; 
(3) the absorption column density changes with time.

In the orientation-based `unified schemes' for radio sources,
flat-spectrum and core-dominated sources are classified as such because
their radio-jet emission is strongly boosted on account of being
viewed at a small angle to the axis of the relativistic jet. 
The cores of more highly-inclined steep-spectrum quasars are 
less strongly boosted, and so their total radio emission is 
dominated by the isotropically emitted radiation from their 
diffuse lobes. Thus, in these models
the prevalence of absorption in steep- rather than flat-spectrum sources, 
and its decrease with $R$, would imply that the absorbing material
lies preferentially away from the jet axis, increasing in column
density the greater the inclination angle.  
Similar results were interpreted this way in earlier studies 
(Anderson et al. 1987;
Foltz et al. 1988; Barthel et al. 1997; Richards et al. 1999, 2001). 
The merits of the unified schemes themselves are argued at length
elsewhere (see reviews by Antonucci 1993; Padovani \& Urry 1995). 
However, orientation alone cannot explain the range of sizes
of steep-spectrum quasars, including CSS 
quasars, so another process is required to explain the decrease
in \civ\ absorption column density with increasing radio size. 

If the absorbing material arises in dense quasar 
environments then we would expect to see its signature 
clearly either in studies of the host galaxies or
the cluster environments of quasars. If this were the case, 
steep-spectrum and small sources (CSS quasars) should
inhabit more dense, gas-rich environments than flat-spectrum 
and larger  sources, respectively. However, there
is currently no observational evidence to support this simple view
(O'Dea 1998; Wold et al. 2000; Saikia et al. 2001). 
Instead, it seems that quasars
of all types are found in environments spanning a wide variety 
of richness (Yee \& Green 1987; Gelderman 1996; 
Wold et al. 2000; Palma et al. 2000), 
from average field density to poor groups to rich clusters. 
In this respect the MQS appears to be no different, as 
has been demonstrated in preliminary imaging of MQS fields
with $0.7<z<1.0$ (Bremer \& Baker 
1999; Baker et al. 2001; Bremer, Baker \& Lehnert 2001). 
Although the dispersion
in relative velocities of the \civ\ absorption systems 
($\sigma \sim 1000$ \kms) is comparable  with that of 
galaxies in massive clusters (e.g. Postman, Lubin  
\& Oke 2001), such clusters
should be rare at $z>1$ (e.g. Bahcall \& Fan 1998;
Bode et al. 2001), and so it is unlikely that a trend such as
that seen in Figure \ref{ewl} would be generated by random
cluster galaxies intercepting the line of sight to the quasar.
The host galaxies of all classes of radio-loud AGN 
(radio galaxies, quasars and CSS sources) also appear to be 
virtually indistinguishable in terms of mass, size
and colour (Heckman et al. 1994; Hes, Barthel \& Hoekstra 1995; 
Fanti et al. 2000; de Vries et al. 2000) once the effect of 
beamed emission in quasars
is taken into account.  However, there is evidence that 
some CSS sources are brighter in the far infrared, 
as a consequence of enhanced starburst activity and/or high
dust content (Hes, Barthel \& Hoekstra 1995).

Theoretically, arguments for the pressure confinement of radio lobes 
imply environmental density must affect source size to some extent
(e.g. Barthel \& Arnaud 1996). Asymmetric sources
provide a clear illustration ---  the shorter lobe tends to be associated with 
higher galaxy densities (Bremer et al. 2001; Barr, Bremer, Baker 2001), 
higher radio depolarisation 
(Garrington et al. 1988; Garrington \& Conway 1991; Ishwara-Chandra 
et al. 1998) and radio fluxes, and brighter extended line emission 
(McCarthy, van Breugel \& Kapahi 1991), all of which are
indicative of higher densities on the shorter-lobe side. 
Typical length ratios of lobes in asymmetric sources are about 2--3 or 
smaller, 
indicative of density contrasts of up to an order of magnitude between 
the two sides. However, to constrict sources to scales of tens rather than
hundreds of kpc would require overdensities of several orders of 
magnitude, which is not seen at any other wavelength. 
There is evidence that some CSSs do inhabit moderately gas-rich 
environments, mostly through radio polarisation 
and  H\,{\sc i} 21-cm absorption measurements on sub-kpc scales
(Akujor \& Garrington 1995; van Gorkom et al. 1989; 
Conway 1996; Peck et al. 1999, 2000; Vermeulen 2001), but it is not clear
whether these environments are exceptional.
There is also no evidence that CSS jets are slowed appreciably, as
their  hotspot advance speeds appear normal ($v\sim 0.2c$)
(Taylor et al. 2000). Thus most CSS jets are unlikely to be 
confined permanently by ultra-dense gas, but the range of 
environments that quasars inhabit must introduce 
variations of factors of up to a few in absolute size. Other
factors, such as jet power, will also introduce scatter 
(de Silva et al., in preparation). Some CSS sources 
might also be short-lived. 

If environmental density alone cannot account for the full range of 
source sizes, then age must be important.  To explain the
results of this paper, the absorption column density 
must decrease over the lifetime of an expanding radio source. 
 Recent radio VLBI observations (Readhead et al. 1996; 
Taylor et al. 2000; Peck et al. 2000) and spectral shape analyses
(Murgia et al. 1999; Snellen et al. 2000) do indeed infer ages as young as
$10^{2}$--$10^{4}$ years for CSS sources compared with the canonical 
$10^{7}$--$10^{8}$ years for larger radio sources. Processes which 
might clear the absorbing material are discussed below.

\subsection{Dust in the absorbing clouds}

As well as the strong dependence on radio properties, 
Section \ref{sec:dust}
suggests that the absorbers contain dust, which 
systematically reddens the continuum. The strong correlation 
between absorption strength and continuum slope in 
Figure \ref{ewaopt} suggests, somewhat surprisingly, that the  
dust and highly-ionised gas exist in close proximity, perhaps even 
in the same clouds. 

The fact that reddening is occurring 
at all indicates the presence of small dust grains 
in the quasar emission-line regions. Such grains should be 
destroyed easily by sputtering by hot gas (Ferrara et al. 1991), 
shocks (Jones, Tielens \& Hollenbach 1996), collisions and evaporation.  
De Young (1998) assessed in detail the problem of survivability 
of dust grains in the vicinity of powerful radio galaxies, 
where strong shocks and hard radiation are undoubtedly present, 
and found that under most circumstances 
100--2500\AA\ graphite grains should be completely destroyed within 
$10^4$--$10^6$ years. Special conditions are required for grain survival 
up to $\sim 10^7$ years, namely low shock velocities and 
dense clouds, or alternatively (but more unlikely) 
a hot, rarefied medium. Otherwise, dust replenishment is necessary.
Dust may be created by new and extensive star formation,
entrained from ISM clouds or torus material by turbulent flows within
the radio cocoon, or redistributed by galactic winds. 
Thus, over the lifetime of a radio source, 
dust grains will be created and destroyed by many processes
and over many astrophysical cycles. Yet, the net outcome
of the dust recycling is gradual clearing, at least along 
the radio axis. With respect to models for the
emergence of quasars from their dusty cocoons, we can
set an approximate timescale of $\sim10^{5-6}$ years, 
comparable with the lifetime of CSS sources, for the
ionisation cones to be opened up significantly. 

Because the absorption lines in our dataset are strong and mostly 
saturated, we cannot measure accurate gas column densities
from a simple curve-of-growth analysis. The lack of damping
wings in \lya\ puts a rough upper limit at 
$N_{\rm HI} < 10^{20}$\,cm$^{-2}$ for neutral gas in 
the optical absorber, although the hydrogen is likely 
to be highly ionised in the \civ-absorbing regions so the
total gas mass will be higher. 
To produce reddening of $E_{\rm B-V}=0.3$--1.0, 
the range applicable to MQS quasars (see also Baker 1997), 
would require a dust-to-gas ratio comparable with or 
greater than the Galactic value. For example, if we assume the 
mean Galactic ratio of 
$\langle N_{\rm HI}/ E_{\rm B-V} \rangle = 5.2 
		\times 10^{21}$ cm$^{-2}$ mag$^{-1}$
(Shull \& van Steenberg 1985), then we would need 
$N_{\rm HI} \sim 10^{21}$\,cm$^{-2}$ to produce $E_{\rm B-V}=0.3$--1.0. 
Although a more detailed quantitative analysis is required, 
this indicates that the dust content of the absorbers is relatively
high.

A decline similar to that shown in Figure \ref{ewaopt}
of \ewa \civ\ with increasing UV spectral slope
has been recently reported by Vestergaard (2001) for a 
mixed radio-loud and -quiet sample, suggesting that 
the absorbers in radio-quiet quasars also contain dust. 
If these results can indeed be extended to all quasars, 
an immediate consequence of the presence of dust in the
associated absorbers is that absorbed quasars will tend 
to be missed preferentially in optically-selected samples,
especially those selected in the blue and ultraviolet. 
This may help explain the paucity of associated absorbers in 
such samples (e.g. only 10\% of LBQS quasars show associated
absorption, Weymann et al. 1991; Ganguly et al. 2001). 
Radio-quiet quasars are discussed further in Section \ref{sec:rqq}.
As a precedent, Fall \& Pei (1993) showed that intervening 
damped \lya\ absorbers also redden the light from background 
quasars, resulting in a systematic loss of reddened sightlines from 
quasar absorption-line studies. Therefore, we would 
expect that even larger numbers of AGN will be reddened 
and dimmed by {\it their own\/} dusty halos.

We note that the fraction of CSS sources among MRC\,1-Jy quasars
and radio galaxies is approximately the same, $\sim 23$\%
(Kapahi et al. 1998). 
This implies that the inner torus opening angle is not significantly 
different for small and large sources (Saikia et al. 2001), 
but the broad lines in smaller
sources are generally more reddened by dust at radii
outside the broad-line region. 
We note a class of CSS sources which are borderline
quasars/radio galaxies (e.g. MRC\,B1208--277 and two 
similar objects with spectra shown in Baker et al. 1999) where 
the broad lines are weak or non-existent but
the narrow emission lines have higher excitation than expected for normal 
radio galaxies. These may be highly reddened CSS quasars, but their
numbers are too small to significantly affect the quasar fraction
in the  MRC\,1-Jy sample.

\subsection{Redshift}

We note that the results of this paper are all independent of redshift,  
i.e. in a flux-limited sample 
we found no marked differences between the global properties 
or frequency of \civ\ associated absorption in quasars 
at $0.7<z<1.0$ and $1.5<z<3.0$. 
This is contrary to the proposition of Ganguly et al. (2001) that
strong \zaze\ absorption systems are rarer at $z<1.2$ than at 
$1.4<z<2.0$. Ganguly et al. (2001) found associated absorption systems
with \ewa$>1.5$\AA\ in only 1/13 steep-spectrum quasars with $z<1.2$
drawn from the UV-bright sample of the HST Absorption-Line Key Project,
compared with 8/24 steep-spectrum quasars with $1.4<z<2.0$ from the sample
of Foltz et al. (1986). The latter fraction, $\sim 30$\%, is in 
good agreement with the number of absorption systems with 
\ewa$>1.5$\AA\ in both our HST (5/17) and ground-based (8/20) datasets.
In addition, the finding that \civ\ absorbers are just as common
at $z\sim 0.8$ as at $z\sim 2$ may have implications for the
interpretation of associated \lya\ absorbers in high-redshift radio galaxies
(Barthel \& Miley 1988; R\"ottgering et al. 1995; 
van Ojik et al. 1997; De Breuck et al. 2000). 
Interestingly, the associated absorption in high-redshift radio galaxies
is found to be most common in sources less than 50\,kpc in size, raising the
idea that some high-redshift radio galaxies are essentially similar to
CSS quasars.

Redshift-independence indicates that the absorbing material
arises in structures which are not evolving significantly with cosmic time.
This point argues against many galactic origins for the absorbing clouds, 
for example in surrounding clusters, star-forming and dwarf galaxies, which 
should show strong evolution between these epochs.  
There remain two plausible explanations: either the material
is linked directly to processes within the active nucleus, or 
we see the absorbers only as a transient property of a 
galaxy system viewed at one particular stage in its evolution.
In the latter case, the phase must be seen
commonly alongside the AGN phenomenon, even if it is not intrinsic to it.

Processes intrinsic to the AGN include winds or outflows 
from the quasar nucleus, which should be governed by basic 
properties of the AGN central engine (black-hole mass, spin, fuelling rate). 
Otherwise, the absorbers might mark a unique environment
or jet-triggering event, which is essential for the formation
of radio jets. For example, mergers between massive galaxies 
might produce similar outcomes, irrespective of redshift.

Alternatively, competing effects (quasar luminosity, redshift) 
might somehow have contrived to cancel out evolutionary trends in our 
flux-limited sample. The ranges of radio power at 408\,MHz 
(expressed in W\,Hz$^{-1}$) sampled
for the low- and high-redshift MQS samples are 
$27.3 < \log P_{408} < 28.2$
(median $ \log P_{408}=27.7$) and
$28.1 < \log P_{408} < 29.0$ 
(median $ \log P_{408}=28.5$), 
respectively (Kapahi et al. 1998). So, the high-redshift sample
is about a factor of 6 more luminous in the radio on average.
In terms of optical B-band luminosity, the difference is
less pronounced and the scatter greater, presumably because 
of the effects of dust. B-band optical luminosities 
(again expressed in W\,Hz$^{-1}$) span 
$22.3 < \log L_{\nu} < 24.6$
(median $\log L_{\nu}= 23.7$) and 
$23.6 < \log L_{\nu} < 25.2$ 
(median $\log L_{\nu}= 24.1$)
for the low- and high-redshift quasar samples,
respectively.
Measurements of larger numbers of absorbers over a wider range in redshift 
and luminosity would be needed to distinguish between these scenarios.

\subsection{Origin of absorbing clouds}

The most fundamental question remains the origin of the 
absorbing gas. Where distances to the clouds can be estimated 
from excited-state fine-structure absorption lines (which necessarily 
require the presence of dense clouds far from  the nuclear 
radiation) they are large --- of order tens of kpc (Hamann et al. 2001). 
This would put the clouds in or near the extended narrow 
emission-line regions 
of quasars, where we already see large quantities of ionised gas 
(Lehnert et al. 1992, Bremer et al. 1992)
and dust (Andreani, Franceschini, Granato 1999; 
Polletta et al. 2000; Haas et al. 2000; Vernet et al. 2001). 

Typical line widths measured for extended (narrow) emission-lines are 
of order 1000\,\kms\ (Lehnert \& Becker 1998;
Heckman et al. 1991; van Oijk et al. 1997; Baum \& McCarthy 2000),
which is comparable with the range of absorption velocities seen here,
and suggests that the gas is turbulent. 
The origin of the large velocities is uncertain, as it is difficult
to disentangle the effects due to the radio jet from those 
arising simply in the surrounding potential well 
(van Oijk et al. 1997; Baum \& McCarthy 2000). 
Direct interactions between
the radio jet and \lya\ clouds have also been noted (Barthel \& Miley 1988;
van Ojik et al. 1997; Lehnert et al. 1999). 
In several high-redshift radio galaxies \lya\ absorption has been resolved 
spatially out to $\sim100$\,kpc, where it is strongest in radio galaxies 
with sizes less than $50$ kpc and knotty radio structures 
(Barthel \& Miley 1988; R\"ottgering et al. 1995; van Ojik et al. 1997;
De Breuck et al. 2000). Absorption by dense H\,{\sc i} gas on compact 
scales is also seen in radio observations towards many quasars 
(Vermeulen 2001; Morganti et al. 2001), especially
compact sources. Although the amount and location
of the H\,{\sc i} gas are still being assessed, it may be anisotropic
(Morganti et al. 2001) and  jet-cloud interactions have been noted 
in a number of cases (Oosterloo et al. 2000). Absorbing
gas with equivalent $N_{\rm H} \sim 10^{21-24}$\,cm$^{-2}$
has also been inferred from soft X-ray absorption 
seen towards many quasars (Elvis et al. 1994; 
Reynolds 1997; Gallagher et al. 2001).
Thus, the extended emission-line regions of quasars 
are known to contain enough gaseous material to account for the
\civ\ absorption and its kinematics, although 
the exact location of the absorbing clouds 
is still an open question. 

As to its origin, dusty, metal-enriched clouds of 
H\,{\sc i} gas could conceivably 
be relics of a merger with a gas-rich galaxy, and remnants of
star-formation triggered by the merger. However, mergers are
efficient only at low relative velocities, so they would 
predict very low velocity dispersions for the absorbing material, 
contrary to observations.  Alternatively, 
cooling flows might be responsible for depositing the gas
(Bremer, Fabian \& Crawford 1997; Nulsen \& Fabian 2000), although 
quasars are not necessarily in rich clusters, and the
large dust content may be difficult to achieve in a cooling flow model.  
Some AGN halos have kinematics which are inconsistent with 
either scenario (Shopbell, Veilleux \& Bland-Hawthorn 1999). 
Otherwise, the gas may simply arise in the potential wells of 
gas-rich galaxy groups and clusters, or protoclusters, 
a suggestion supported by the
discovery of companion star-forming galaxies around some high-redshift 
radio galaxies and quasars 
(Bremer, Fabian \& Crawford 1997; Pentericci et al. 2000; 
Ivison et al. 2000; Baum \& McCarthy 2000).

An alternative hypothesis is that the 
absorbing clouds are embedded in or ejected by nuclear outflows
or galactic winds. 
Interestingly, Heckman et al. (2000) have reported a 
correlation analogous to Figure \ref{ewaopt} in observations 
of starburst galaxies, namely between the  spectral slope 
of the UV continuum and the equivalent width of Na\,D 
absorption from the dusty superwind. Starburst superwinds
typically cover a much smaller range of velocities than seen
in quasar absorbers, but it is interesting 
to ask whether the quasar absorbers represent an 
extreme case of the galactic wind phenomenon. 
Chen, Lanzetta \& Webb (2001) propose
that $\sim 100$\,kpc \civ-absorbing halos may be common around 
even normal galaxies, presumably enriched by galactic wind activity.
Radio H\,{\sc i} measurements and sensitive optical narrow-band imaging
will be crucial for disentangling the origins of the gas. 

The results of this paper provide additional clues to the origin 
of the absorbing clouds. 
The anticorrelation of absorption strength with source size, if it 
can be interpreted in terms of age, implies that radio jets are 
triggered in an environment rich in gas and dust. Mergers have long
been thought likely causes of AGN activity, as they will efficiently
funnel gas onto the galaxy nucleus. In addition, major mergers
are expected to cause massive starbursts, providing a natural 
explanation for the origin of the dusty material.
Thus, starbursts may have a symbiotic relationship with AGN,
(Gonz\'alez-Delgado, Heckman \& Leitherer 2001; 
Barthel 2001; Canalizo \& Stockton 2000). 
A similar picture has been put forward
by Sanders et al. (1988) where ultra-luminous infrared galaxies 
are proposed as the dusty progenitors of optical
quasars. However, in the
radio-loud case we have been able to demonstrate a natural age 
sequence. We discuss this further in Section \ref{sec:pic}.

Another alternative is that the absorbers originate in
AGN-driven nuclear outflows (Elvis 2000; de Kool 1997; 
Proga, Stone \& Kallman 2000). However, in these models 
the decrease in absorbing column 
density with radio size may be difficult to explain. 
The main mechanism to decrease the outflow rate is to
steadily decrease the accretion rate of material onto 
the black hole, but this should have other observable 
repercussions such as dimming the nuclear light and 
reducing the jet power in large sources.

\subsection{Relationship with radio-quiet quasars}
\label{sec:rqq}

The results reported here raise further questions about the 
apparent differences between radio-loud and radio-quiet quasars. 
It is widely held that radio-quiet quasars have a lower 
incidence of associated absorption than radio-loud quasars. 
The fraction of radio-quiet quasars with  \zaze\ absorption 
ranges from $\sim 10$--15\% (Foltz et al. 1988; Weymann et al. 1991; 
Ganguly et al. 2001) to $\sim 30$\% (Vestergaard 2001), 
compared with 40--50\% of radio-loud quasars in low-frequency
selected samples, including the MQS. Although these numbers
clearly show that consensus has not yet been reached, it has been
found that narrow associated-absorption systems are more common amongst 
certain types of radio-quiet quasars, notably soft-X-ray weak 
quasars (Brandt, Laor \& Wills 2000; Ganguly et al. 2001)
and perhaps low-luminosity quasars (M\o ller \& Jakobsen 1987; 
but see Foltz et al. 1988; and mixed results in Vestergaard 2001). 
Therefore, it seems likely that sample selection criteria will affect the 
frequency of finding \zaze\ systems, although it remains to
be seen how severely. 

If both radio-loud and radio-quiet quasars have a similar 
distribution of absorbing material then the brightest 
optically-selected quasars should be viewed 
preferentially along clear sightlines to the nucleus
within $\sim20$\degree--30\degree\ of some 
polar axis, avoiding most of the obscuring material. 
The presence of dust in the off-axis material would accentuate 
this natural bias further. Therefore, we would expect to find
more associated absorption systems in samples that include
fainter, redder quasars. If the hypothesis that both classes
are similar is true, then we might predict that the proportion
of radio-quiet quasars showing absorption will ultimately 
rise to match that of the radio-loud quasars. Otherwise, the 
distributions of dust and gas must differ.
The recent preliminary results of Vestergaard (2001) indeed point 
to gross similarities between the global reddening and absorption 
properties of radio-quiet and radio-loud quasars, despite the 
difficulties inherent in matching samples across these two classes.
In addition, Crenshaw et al. (1999) have found a significantly
higher rate of occurrence of narrow associated-absorption systems 
in Seyfert 1 galaxies than was first appreciated, a fraction
of $\sim 50$\% comparable with the radio quasars in this paper. 
We note that the classification of AGN as either radio-loud 
or -quiet is especially difficult for AGN of moderate radio luminosity,
including Seyfert galaxies, where the radio and optical emission 
may be contaminated by starburst activity, emission from the host 
galaxy and relativistic boosting (e.g. Miller, Rawlings \& Saunders 1993; 
Falcke, Sherwood \& Patnaik 1996; Ho \& Peng 2001; and see 
Blundell \& Rawlings 2001).

In the far-infrared, the distribution of dust in radio-loud and 
radio-quiet quasars is still under investigation. Both classes have 
similar far-infrared properties (Andreani et al. 1999; Haas et 
al. 2000; Polletta et al. 2000), 
suggesting they harbour comparable amounts of cool dust,
characteristic of dusty ISM warmed by star-formation.
More work is needed, however, to 
investigate warmer dust components in the two classes, 
where emission reprocessed by the circumnuclear torus dominates. 

If the covering factor of the absorbing clouds declines
with age, then many young, dust-enshrouded 
AGN will be missing from optically-selected samples.
The precise number would depend on the timescale for
clearing the material. Based on the assumption that
ultra-luminous infrared galaxies (ULIRGs) evolve into quasars
(e.g. Sanders et al. 1988), Canalizo \& Stockton (2001) 
estimate that it takes $\sim50$\,Myr for 
a radio-quiet quasar to emerge from its ULIRG phase.
This timescale is longer than the CSS lifetime ($<1$\,Myr) 
discussed in Section \ref{sec:discrad}, suggesting that
dust is dissipated  faster in quasars with radio jets.

In optically-selected quasars, Brandt et al. (2000) reported
a strong correlation between \ewa(\civ) and the optical to X-ray slope,
$\alpha_{\rm ox}$, linking \civ\ and soft X-ray absorption. 
We are not able to see a similar correlation in our sample for the
few quasars we have detected in the ROSAT All 
Sky Survey (Baker, Hunstead \& Brinkmann 1995), but this is
not surprising given the large number of non-detections.
However, Baker et al. (1995) did note that CSS quasars as a group
were detected much less often in the RASS, which may indicate 
X-ray absorption or intrinsic weakness (see also O'Dea 1998). 
Obviously more X-ray data are needed for the MQS.  

As for broad absorption lines (BALs), 
which favour radio-quiet quasars and originate much closer to the AGN, 
it is still an open question why they
are seen so rarely in radio-loud objects. 
We note that there is now considerable evidence for dust
in BAL systems (Sprayberry \& Foltz 1992; 
Turnshek et al. 1994; Egami et al. 1996; Brotherton et al. 1998).

\subsection{Consistent overall picture}
\label{sec:pic}

Given the redshift-independence and clearing of absorbing 
material with time, and especially the evidence that radio activity 
can be recurrent, the presence of dusty absorbing gas {\it must\/} 
be intimately related to the onset of the radio activity, 
rather than just the host galaxy or cosmic time. 
Such a build-up of dusty material  would occur naturally following 
a large burst of star-formation, either during or following a merger 
with a gas-rich galaxy. 

Over the source's lifetime, the enshrouding material is gradually
cleared, primarily along the jet axis. The bow-shock of the radio source
will be particularly efficient in compressing and heating the 
clouds and shattering dust grains as it passes. The absence 
of any absorption close to the jet axis itself 
(in core-dominated, flat-spectrum quasars) suggests that the 
direct interaction of the radio beam on the ISM clouds
may be particularly effective in shocking and displacing 
material near the hotspot region, punching a hole in the 
enshrouding gas. 
Away from the jet, 
substantial replenishment of dust is likely to occur
by, for example, turbulent entrainment at the edges 
of the cocoon, evaporation from the torus, extended 
star-formation or galactic winds. Dust may be deposited
in disks, or fed back into the torus. Thus
the torus must comprise recycled ISM material, and vice versa. 
Turbulence inside the lobe cavity may also shred clouds, 
exposing the large surface area to ionizing radiation 
and producing aligned optical emission with high velocity dispersion 
(Bremer et al. 1997; Best, Longair \& R\"ottgering 1996).

The basic test for this picture is that CSS sources
should show more evidence of recent star-formation
than larger sources. There is so far mixed evidence that
CSS quasars are brighter in the far-infrared (Hes et al. 1995).
On a more detailed level, one might compare the 
estimated ages of starburst knots and the radio source.
At least in one case, the CSS quasar 3C48 (Canalizo \& Stockton 
2000), the ages are found to be consistent. 
Extending this technique to higher redshifts, 
however, is more difficult. Estimating ages for radio sources
is also a fine art; a new practical method applied to the
present sample will be presented by de Silva et al. (2001, in preparation). 

Finally, we note that apart from the merger/starburst hypothesis,
the only plausible alternative is that an AGN-driven wind model
might be contrived in which the mass flow declines markedly
$10^{5}$ years after the onset of radio activity.

\section{Conclusions}

We present a study of \civ\ associated absorption in a 
highly-complete, homogeneous sample of radio-loud quasars. 
The main results are:

\begin{itemize}

\item We confirm that absorption is more common in 
steep-spectrum and lobe-dominated quasars, such that the absorbing
material lies away from the jet axis in
the orientation-dependent unified models.

\item The strength of \civ\ absorption decreases
with increasing radio-source size. If we assume that the larger 
sources are older than the smaller (CSS) ones, then we can
attribute the decrease in column density to the
growth of the radio-source envelope through the ISM of the host galaxy.

\item From the correlation of \civ\ absorption strength with optical
spectral slope, we conclude that considerable amounts of dust are 
associated with the absorbing clouds. Consequently we predict that
absorbed quasars will be missed preferentially in optically-selected
samples, provided similar schemes apply.

\item We find no evidence for changes in the frequency
or strength of the absorbers with redshift from $z\sim 0.7$ to $z\sim 3$.
This lack of cosmic evolution indicates the
absorbers are unaffected by gross galaxy evolution, 
rather they signal a transient phase which is
related specifically to the AGN activity.

\item The combination of these results requires that radio
sources are triggered in gas-rich, dusty galaxies, such as
those immediately following
a starburst, and the dust and gas dissipates over the
lifetime of the radio source.
Thus, the ionization cones may open up with increasing radio-source age.

\end{itemize}

\bigskip
\acknowledgements

The referee is thanked for a detailed reading of the manuscript
and useful comments. 
We thank the staff at the Anglo-Australian Telescope and at 
ESO, La Silla and the VLT for their help during observing runs. 
We also thank Tara Murphy for her work on the HST data.
JCB acknowledges support 
which was provided by NASA through Hubble Fellowship grant 
\#HF-01103.01-98A from the Space Telescope Science Institute, 
which is operated by the Association of Universities
for Research in Astronomy, Inc., under NASA contract NAS5-26555,
and also PPARC.
RWH acknowledges funding from the Australian Research Council.

%
%

\newpage
\begin{figure*}[p]
\centerline{\psfig{file=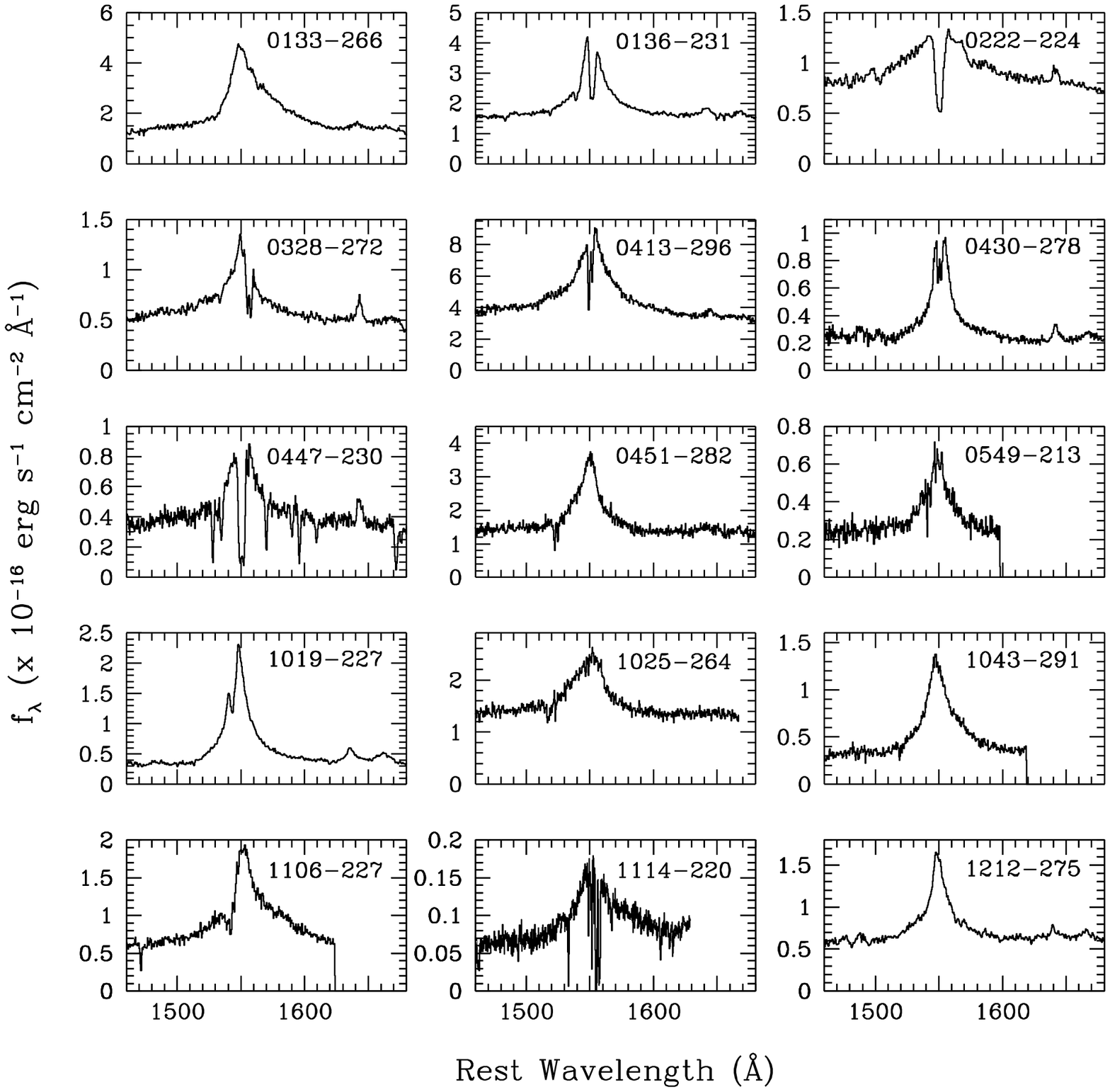,bbllx=
20pt,bblly=140pt,bburx=582pt,bbury=700pt,height=20cm} }
\caption[e]{\footnotesize
Spectra of \civ\ region (1400--1700\AA) for $z>1.5$ quasars observed
from the ground. 
}
\label{grdspectra}
\end{figure*} 

\newpage
\setcounter{figure}{0}
\begin{figure*}[p]
\centerline{\psfig{file=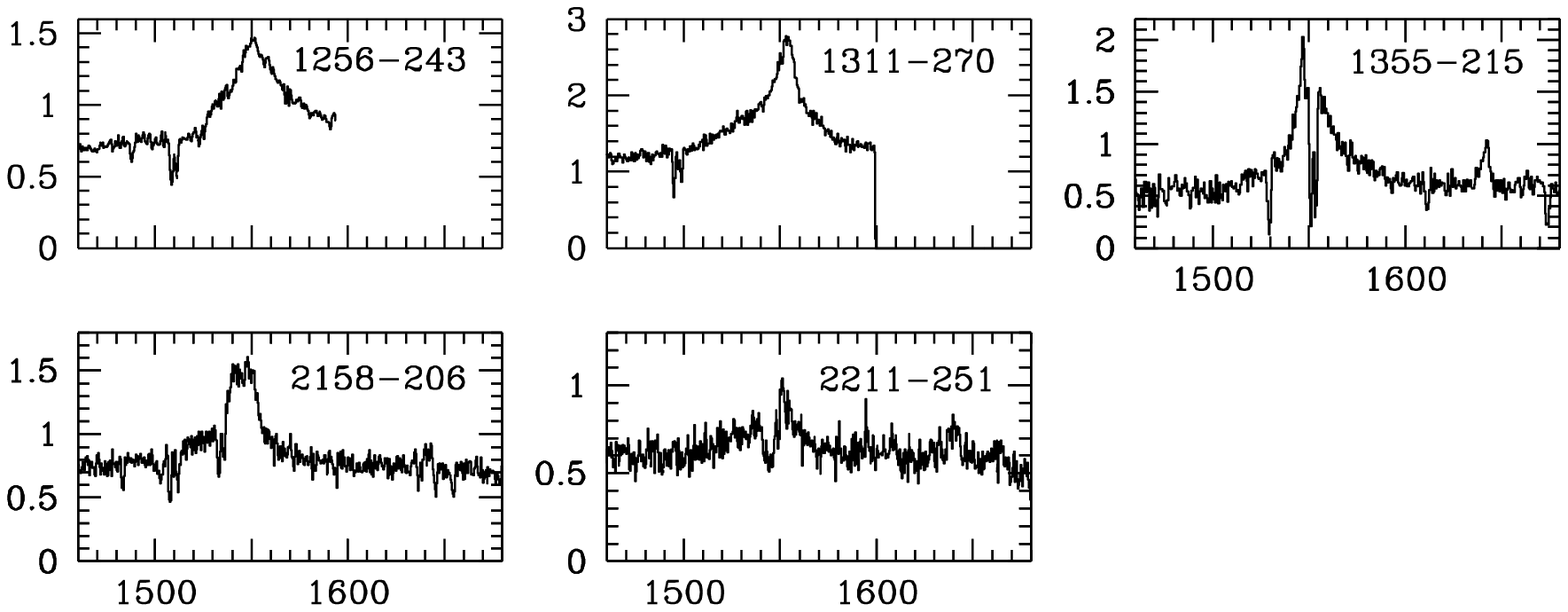,bbllx=
20pt,bblly=459pt,bburx=582pt,bbury=706pt,height=8cm} }
\caption[e]{\footnotesize
{\it Continued}. 
Spectra of \civ\ region (1400--1700\AA) for $z>1.5$ quasars observed
from the ground. 
}
\end{figure*} 

\newpage

\begin{figure*}[p]
\centerline{\psfig{file=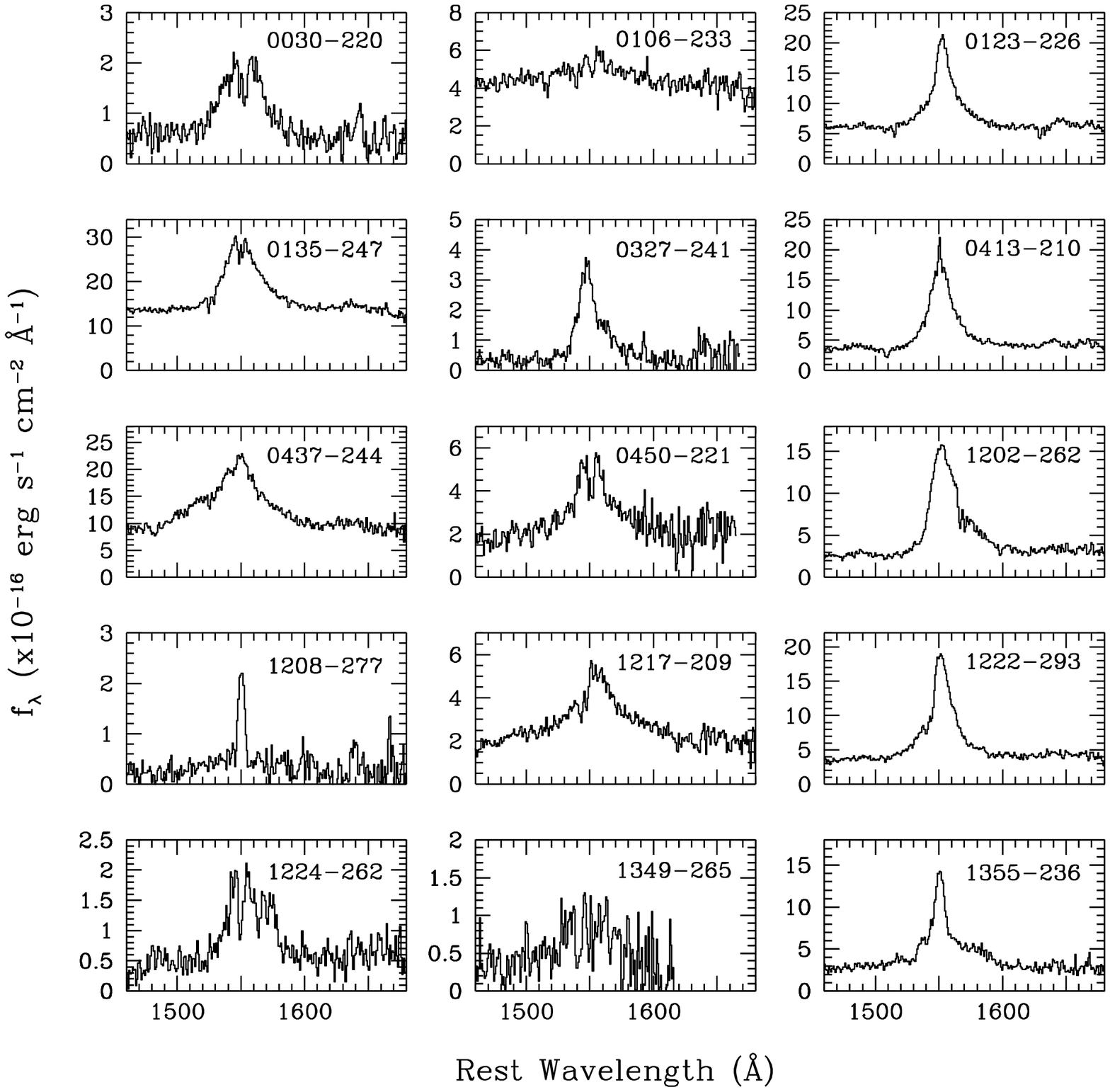,bbllx=
20pt,bblly=140pt,bburx=582pt,bbury=700pt,height=20cm} }
\caption[e]{\footnotesize
Spectra of \civ\ region (1400--1700\AA) for $0.7<z<1.0$ quasars observed
with HST. 
}
\label{hstspectra}
\end{figure*} 

\newpage
\setcounter{figure}{1}
\begin{figure*}[p]
\centerline{\psfig{file=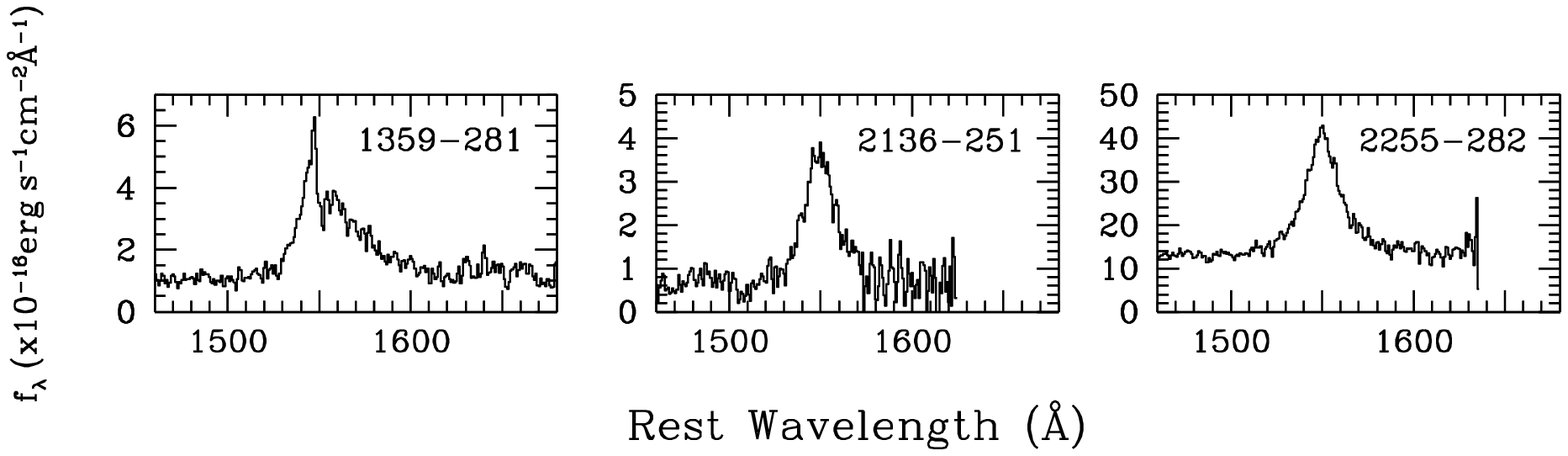,bbllx=
20pt,bblly=570pt,bburx=582pt,bbury=720pt,height=5cm} }
\caption[e]{\footnotesize
{\it Continued}. 
Spectra of \civ\ region (1400--1700\AA) for $0.7<z<1.0$ quasars observed
with HST. 
}
\end{figure*}

\newpage

\begin{figure*}
\centerline{\psfig{file=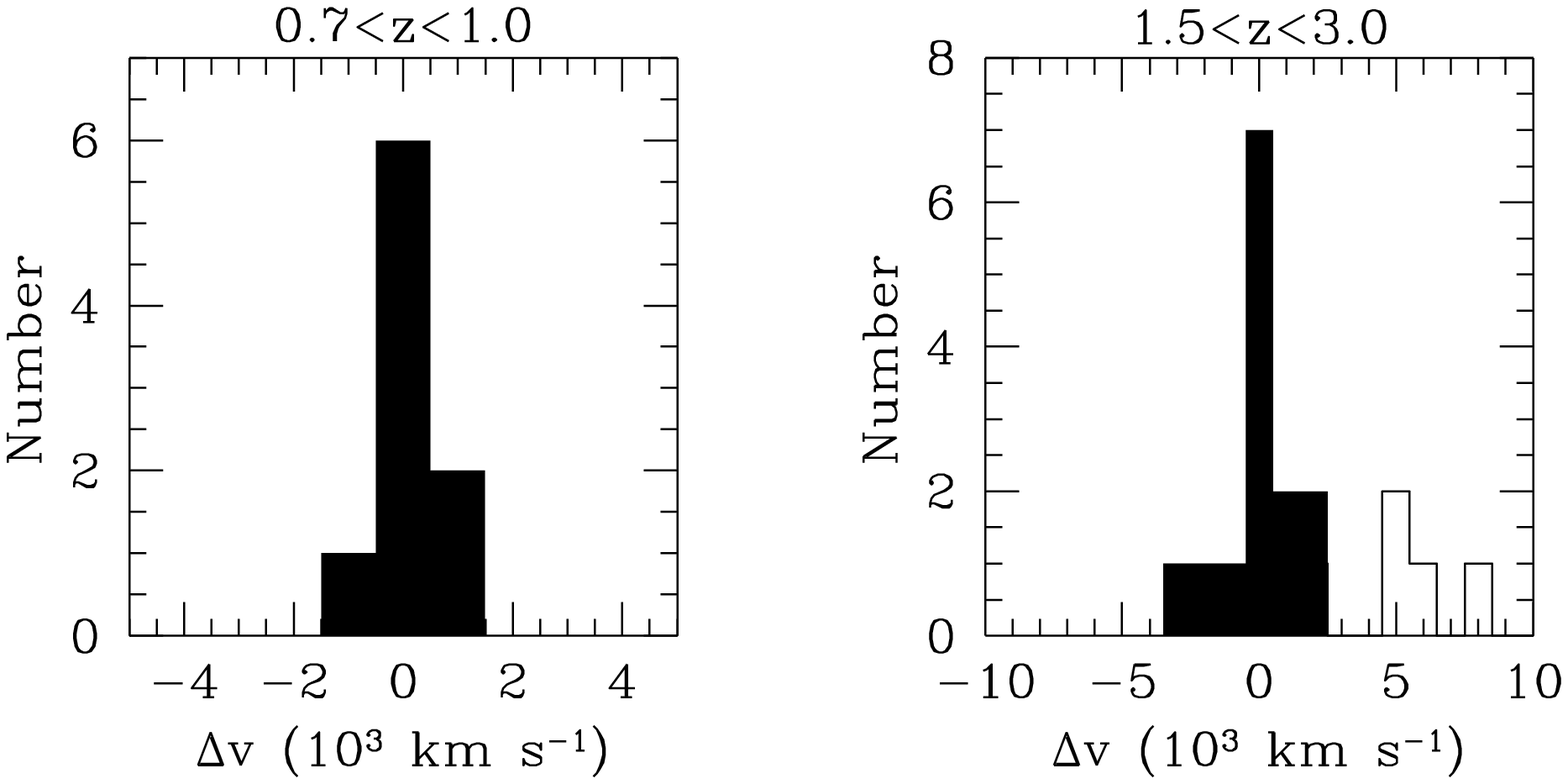,bbllx=
30pt,bblly=150pt,bburx=585pt,bbury=435pt,height=8cm} 
}
\caption[e]{\footnotesize
Histograms of velocities of the \civ\gl1548 absorption line
(or \gl1550 blend) measured relative 
to the broad \civ\gl1550 emission-line peak 
($\Delta v = c (z_e-z_a) / (1+z_e)$). 
Quasars with $0.7<z<1.0$ (HST observations) are shown on the left,
those with $1.5<z<3.0$ (ground-based observations) on the right. 
Absorption lines with positive $\Delta v$ are blueshifted with respect to
the emission redshift. Systems with $|\Delta v| < 5000$ \kms\ are shaded.
}
\label{vels}
\end{figure*}

\newpage

\begin{figure*}
\centerline{\psfig{file=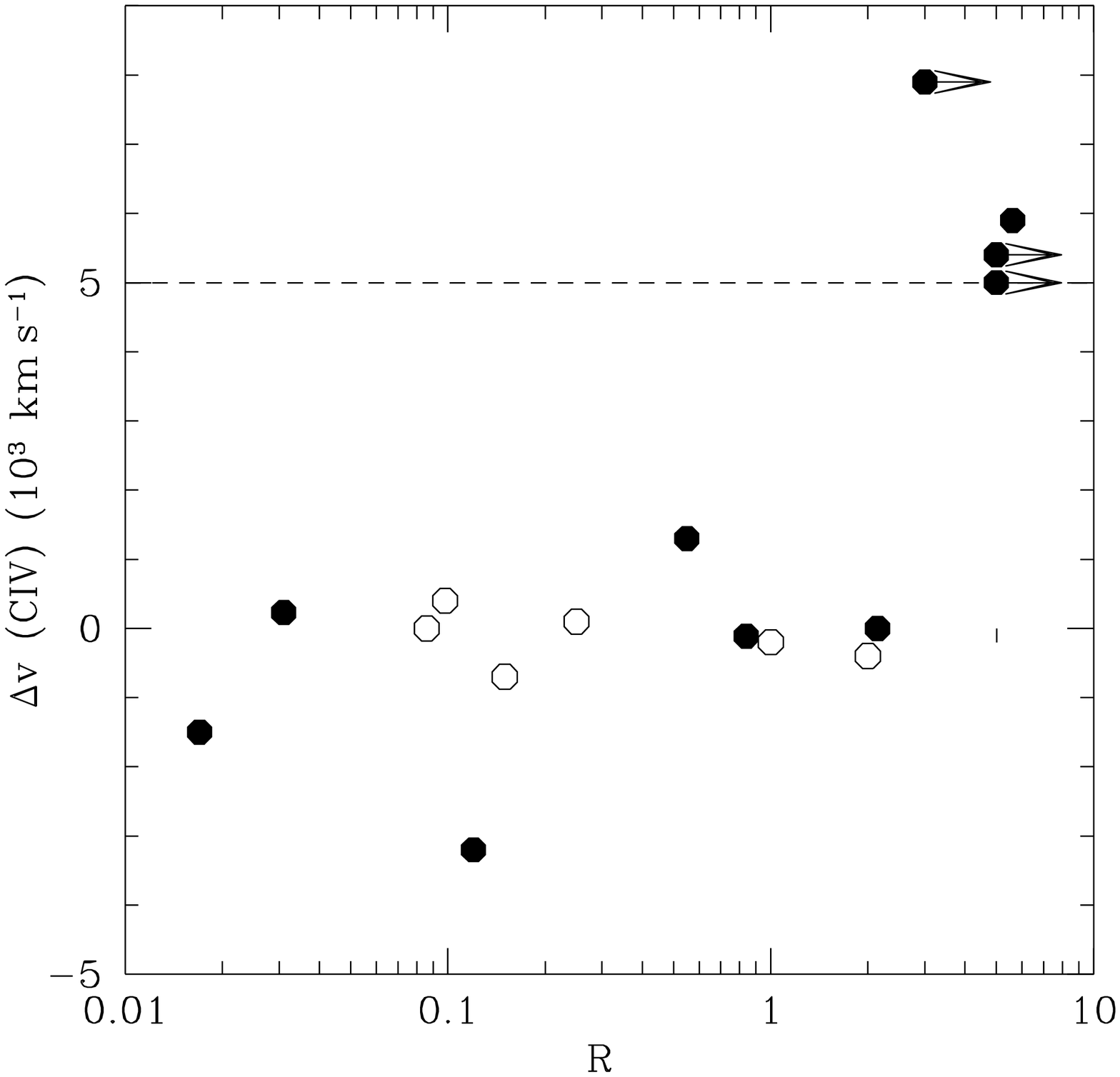,bbllx=
20pt,bblly=140pt,bburx=585pt,bbury=705pt,height=14cm} }
\caption[e]{\footnotesize
Relative velocity of \civ\ absorption as a function of radio
core-to-lobe ratio, $R$ (measured at 10-GHz in the restframe),
for MQS quasars with $1.5<z<3.0$ (filled symbols)
and $0.7<z<1.0$ (open symbols). Note all 4 quasars with 
$\Delta v \geq 5000$\,\kms\ absorption are very highly 
core dominated ($R > 3$). 
}
\label{rvel}
\end{figure*} 

\begin{figure*}
\centerline{\psfig{file=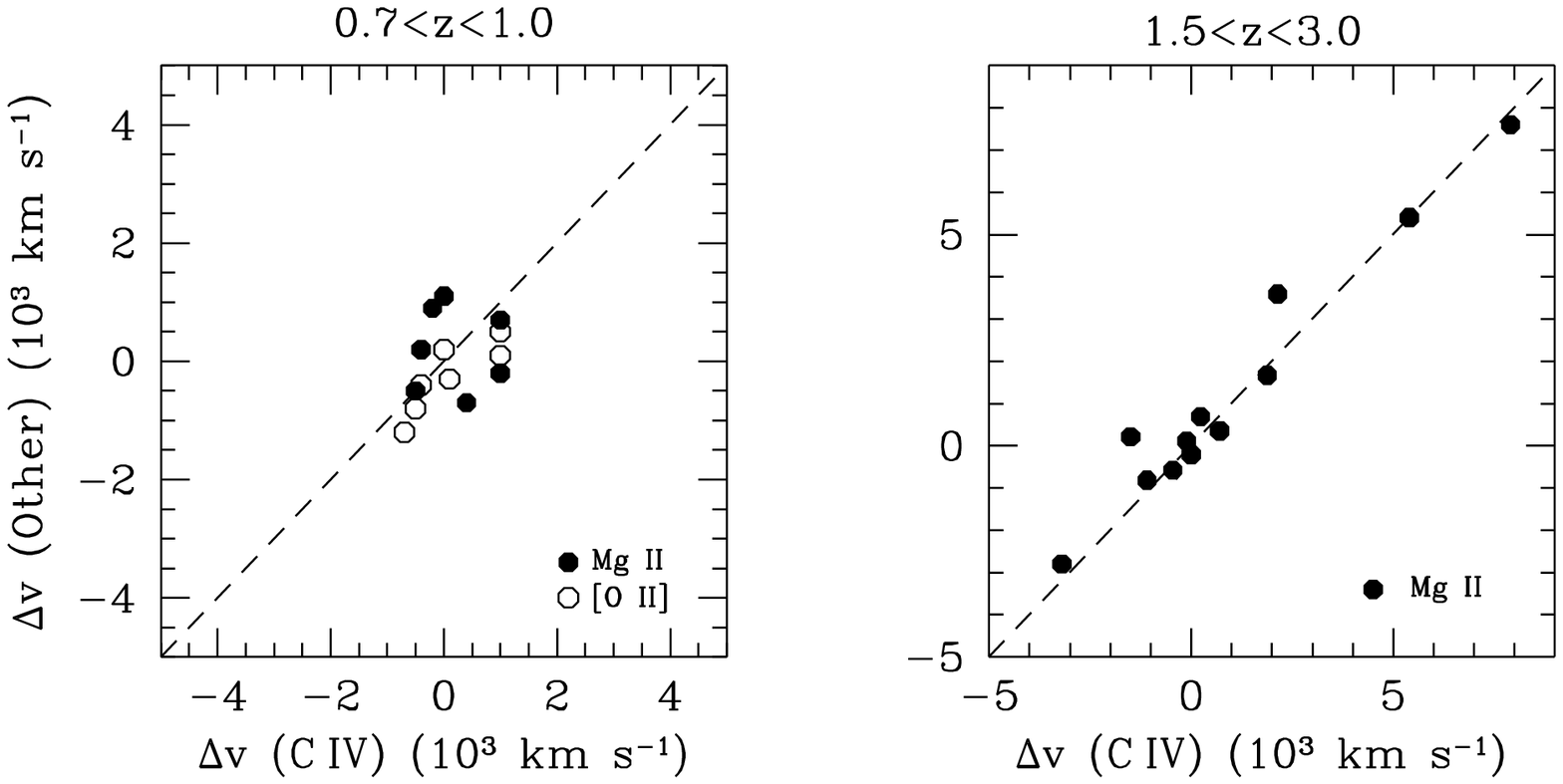,bbllx=
20pt,bblly=150pt,bburx=585pt,bbury=435pt,height=8cm} }
\caption[e]{\footnotesize
Comparison of velocities of the \civ\gl1548 absorption
line (or \gl1550 blend) measured relative to the 
emission-line redshifts defined by the broad 
\civ\gl1550 and \mgii\gl2800 (filled circles)
or narrow \oii\gl3727 (open circles) emission lines for low-
({\it left\/}) and high- ({\it right\/}) 
redshift datasets. Measurements relative
to \oii\ are for $0.7<z<1.0$ quasars only. If there are no
systematic shifts, then the points should scatter about 
the dashed lines, within the measurement uncertainties (see text). 

}
\label{velcomp}
\end{figure*}

\begin{figure*}
\centerline{\psfig{file=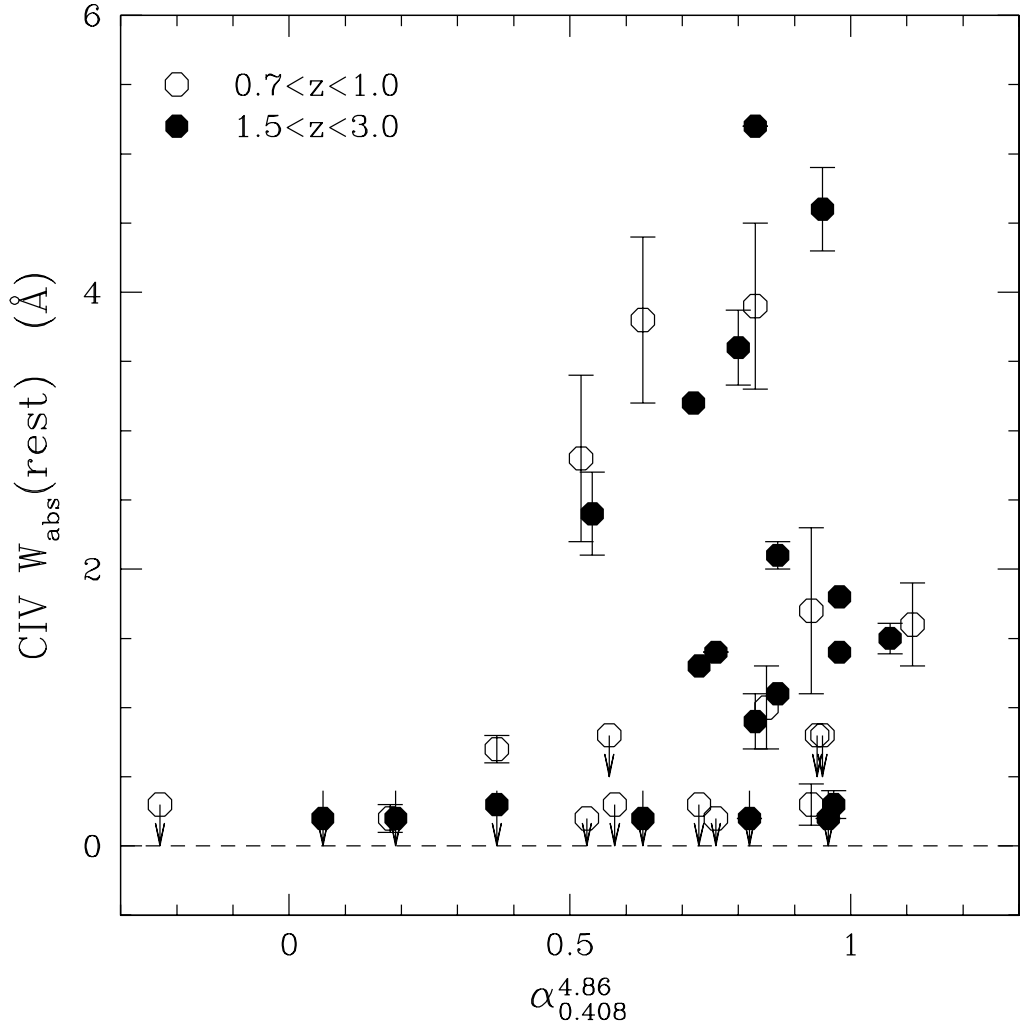,bbllx=
20pt,bblly=140pt,bburx=575pt,bbury=705pt,height=14cm} }
\caption[e]{\footnotesize
Equivalent width, \ewa, of \civ\gl\gl1548,1550 
absorption doublet ($\Delta v < 5000$\,\kms) as a function of 
radio spectral index, $\alpha_{0.408}^{4.86}$ 
($S_{\nu} \propto \nu^{-\alpha}$; measured between 408\,MHz and
4.86\,GHz), for MQS
quasars. Quasars with $0.7<z<1.0$ are plotted as open circles, 
those with $1.5<z<3.0$ are plotted with filled symbols.
Arrows indicate upper limits. 
}
\label{bothewarad}
\end{figure*} 

\newpage

\begin{figure*}
\centerline{\psfig{file=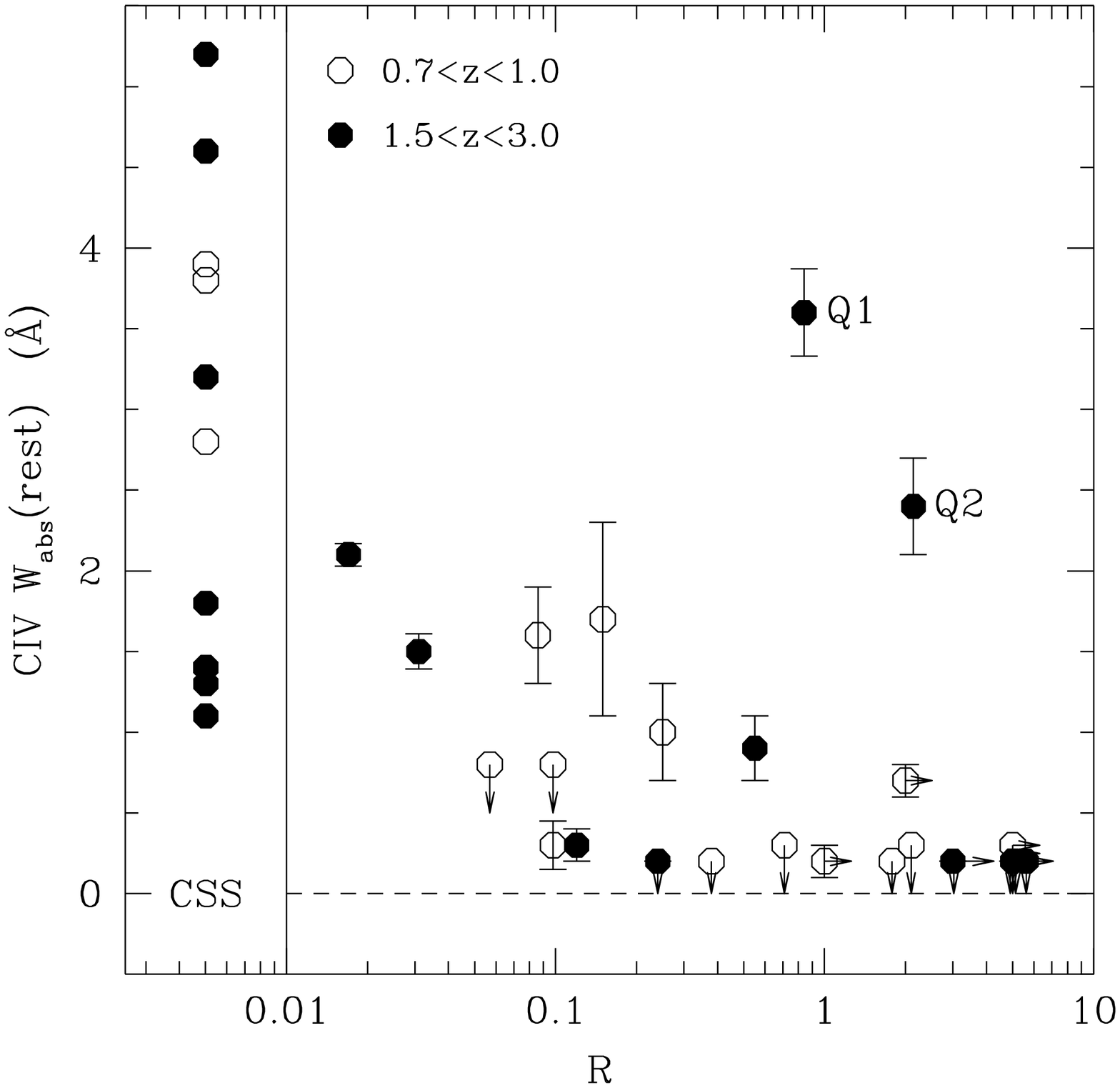,bbllx=
20pt,bblly=140pt,bburx=575pt,bbury=705pt,height=14cm} }
\caption[e]{\footnotesize
Equivalent width of \civ\ absorption as a function of 
radio core-to-lobe ratio, $R$. Quasars
with $0.7<z<1.0$ (HST data) are plotted as open circles, 
those with $1.5<z<3.0$ are plotted with filled symbols.
Limits are shown with arrows. Two outliers 
are labelled (see text), i.e. B1355--215 (Q1) 
and B0136--231 (Q2).
CSSs do not have $R$ measurements and are plotted
separately on the left for comparison. 
}
\label{rew}
\end{figure*} 

\clearpage

\newpage

\begin{figure*}
\centerline{\psfig{file=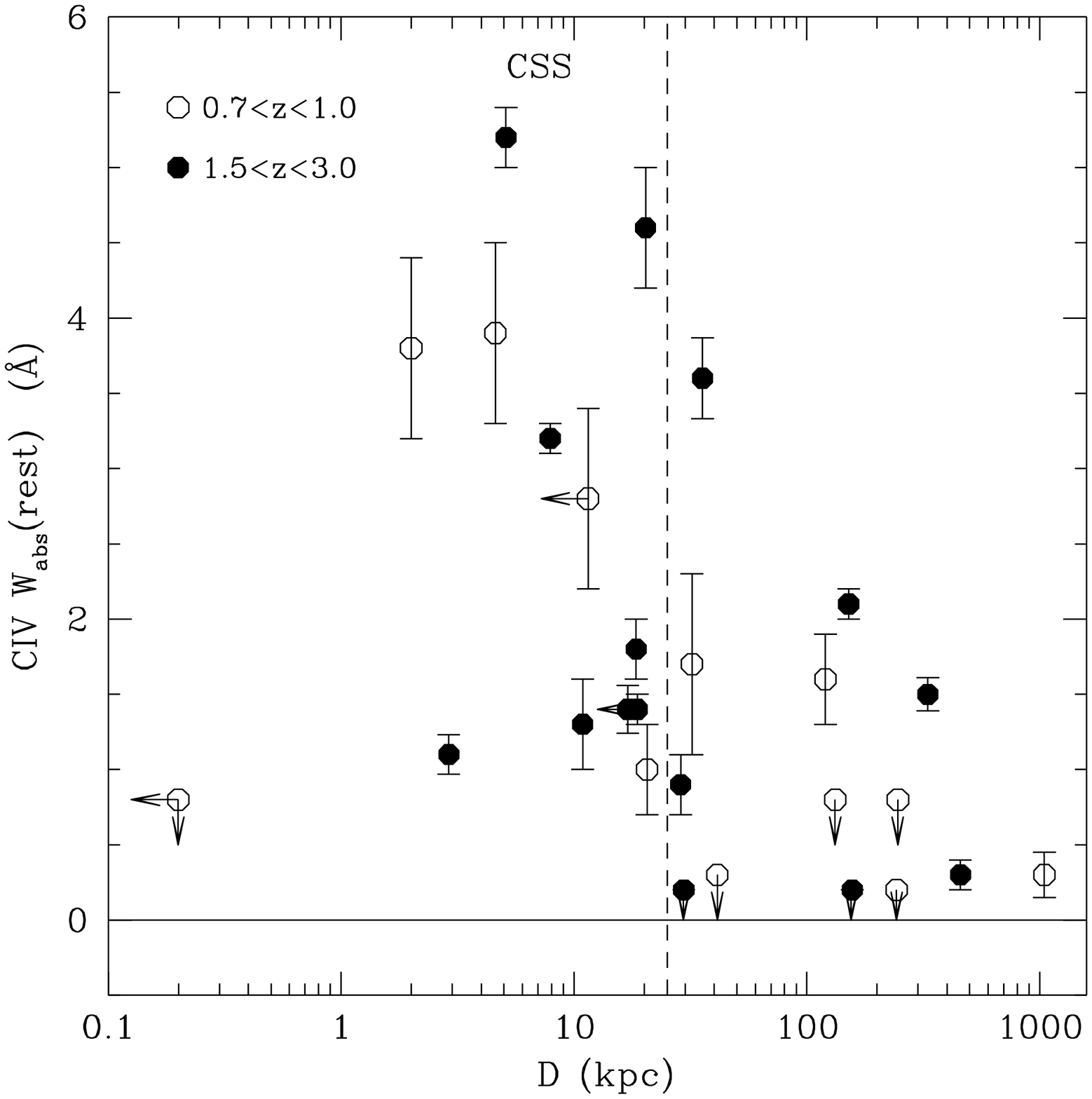,bbllx=
20pt,bblly=140pt,bburx=575pt,bbury=705pt,height=15cm} }
\caption[e]{\footnotesize
Equivalent width of \civ\ absorption as a function of 
radio source size, $D$ (kpc).  
Quasars with $0.7<z<1.0$ are plotted as open circles, 
those with $1.5<z<3.0$ are plotted with filled symbols.
The dotted line at $D=25$ kpc
illustrates the size limit used to define CSSs; 
we note this definition is somewhat arbitrary.
The smallest quasar ($<0.2$ kpc) is GPS source B2136--251.

}
\label{ewl}
\end{figure*}

\begin{figure*}
\centerline{\psfig{file=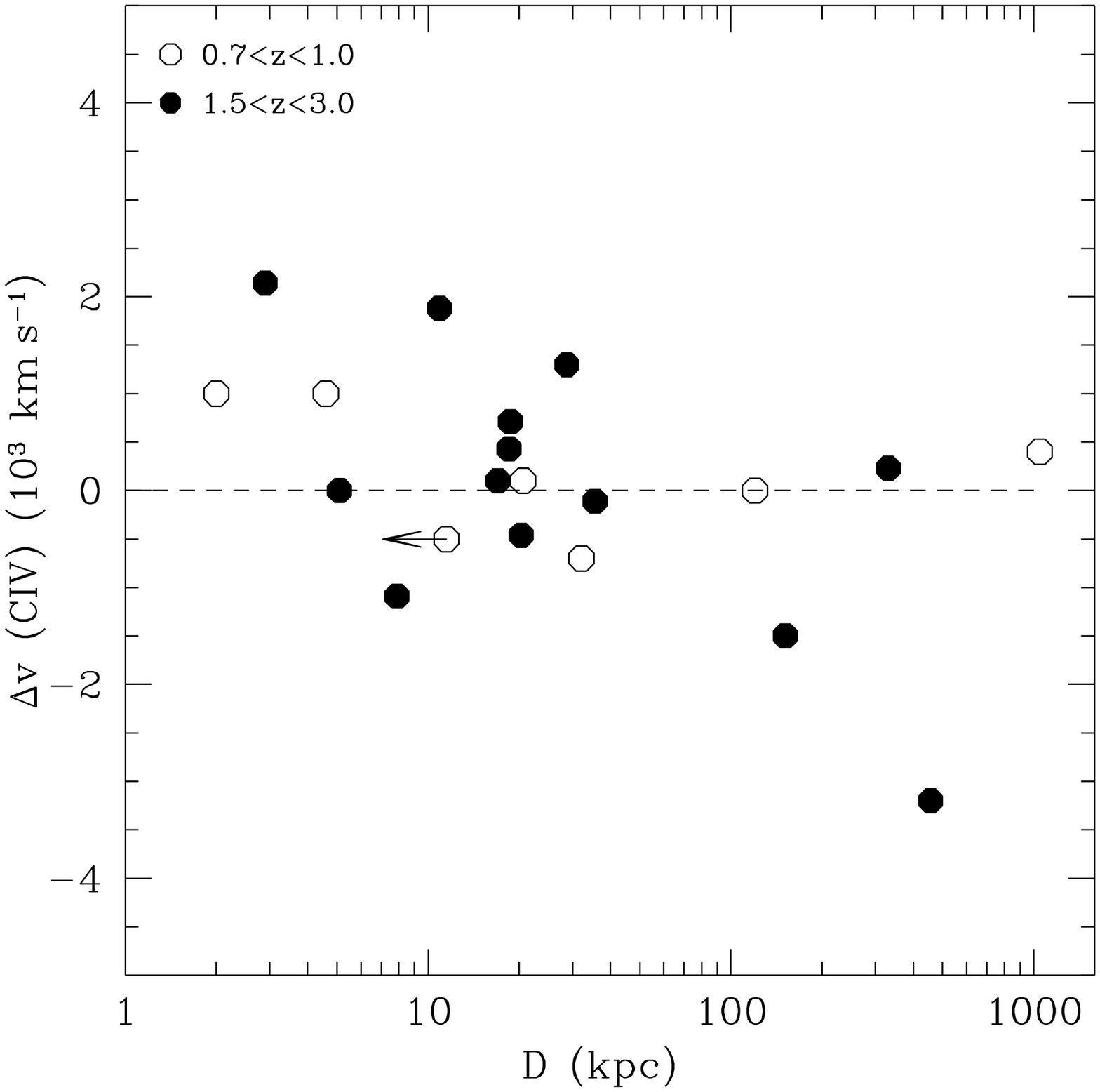,bbllx=
20pt,bblly=140pt,bburx=575pt,bbury=705pt,height=14cm} }
\caption[e]{\footnotesize
Velocity of narrow \civ\ absorption relative to broad \civ\ emission
as a function of radio linear size ($D$). 
Positive absorption velocities are blueshifted 
with respect to the emission redshift.
Quasars with $0.7<z<1.0$ are plotted as open circles, 
those with $1.5<z<3.0$ are plotted with filled symbols.
Velocity uncertainties are roughly $\pm 600$ \kms.
}
\label{lvel}
\end{figure*}

\begin{figure*}
\centerline{\psfig{file=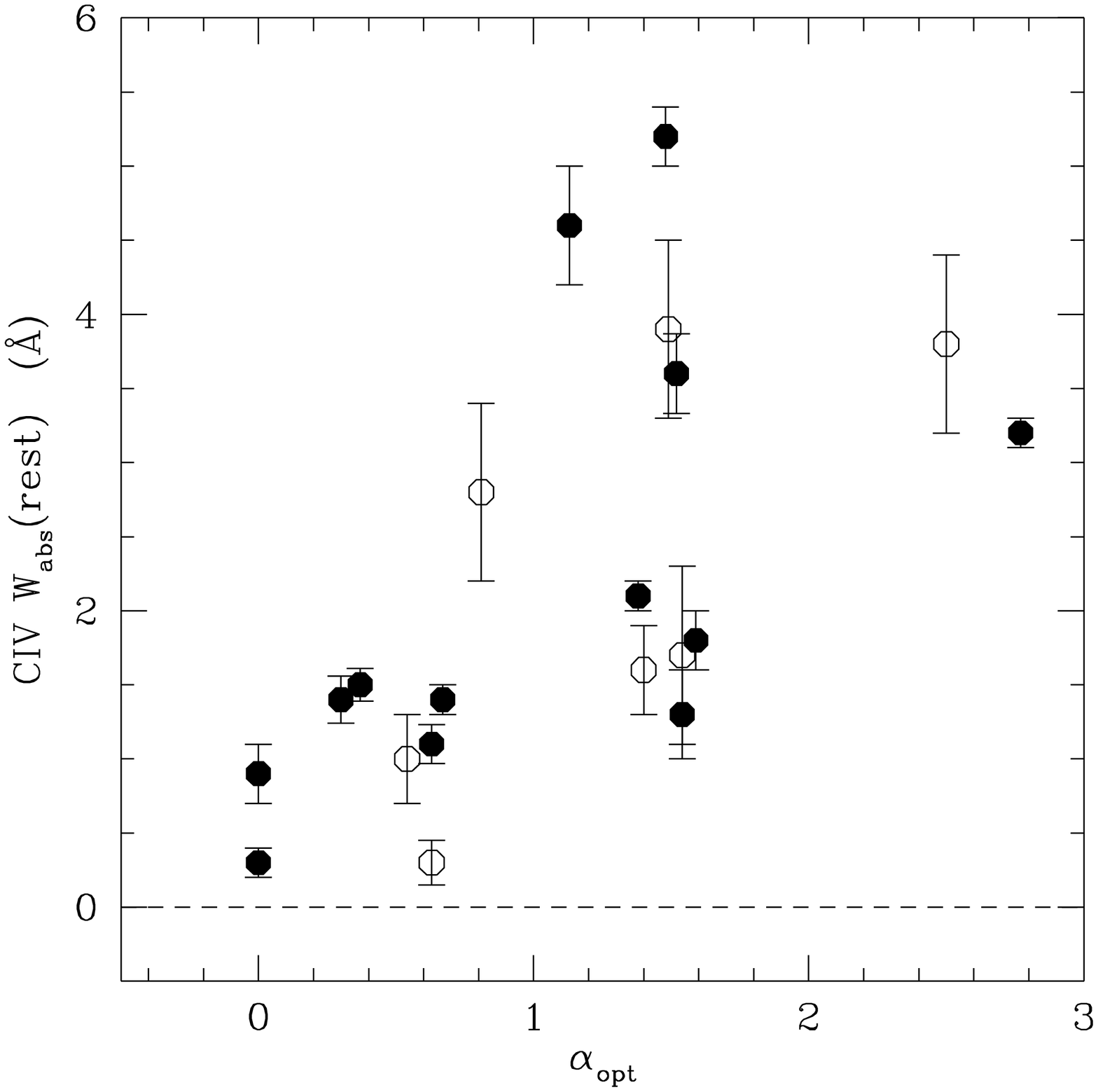,bbllx=
20pt,bblly=140pt,bburx=575pt,bbury=705pt,height=14cm} }
\caption[e]{\footnotesize
Equivalent width of \civ\ absorption doublet as a function of 
optical spectral index, $alpha_{\rm opt}$ 
($S_{\nu} \propto \nu^{-\alpha}$;
measured between 4000 and 10000\AA)
for all quasars with detected absorption. 
Quasars with $0.7<z<1.0$ are plotted as open circles, 
those with $1.5<z<3.0$ are plotted with filled symbols.
}
\label{ewaopt}
\end{figure*}

%

\newpage
\clearpage


\renewcommand{\arraystretch}{.6}
\begin{table*}
\begin{minipage}{17cm}
\caption[]{Observational data for all MQS quasars with $1.5<z<3.0$}
\label{hightab}
\begin{tabular}{@{}cclcrccclcl}
\\
\tableline
\footnotesize
MRC Quasar & $\langle z_e \rangle$ & $b_{\rm J}$ & $R$ & $D$ (kpc) &
$\alpha_{\rm opt}$ & $\alpha_{0.408}^{4.86}$ 
& Tel & UT Date & Res (\AA) & Notes \\
(1) & (2) & (3) & (4) & (5) & (6) & (7) & (8) & (9) & (10) & (11) \\ 
\tableline
\\
B0133--266 & 1.530 &  19.9  &  0.12   &   456.0  & 0.0 &   0.97  & ESO & 00-FEB-02 & 2.0 & S\\
B0136--231 & 1.895 &  19.7  &  2.14   &   105.9  & 0.5 &   0.54  & ESO & 00-FEB-01 & 2.0 & S\\
B0222--224 & 1.601\tablenotemark{\ast} 
                   &  19.1  &  CSS    &    20.3  &1.13 &   0.95  & ESO & 00-FEB-01 & 2.0 & S\\
           &	&		& & & &				 & AAT & 95-OCT-20 & 1.2 &  \\
B0237--233 & 2.224 &  16.4  &  GPS    &   $<0.1$ &0.64 &   0.05  & AAT & 95-OCT-19 & 1.2 & \lya\ only \\
B0246--231 & 2.914 &  21.4  &  CSS    &   $<1.8$ &1.92 &   0.70  & AAT & 95-OCT-19 & 1.2 & \lya\ only\\
B0328--272 & 1.803 &  18.1  &  0.017  &   151.1  &1.38 &   0.87  & ESO & 00-FEB-01 & 2.0 & S\\
B0413--296 & 1.608\tablenotemark{\ast} 
		   &  18.6  &  0.031  &   330.4  &0.37 &   1.07  & AAT & 97-FEB-09 & 2.4 & S\\
           & & & & & & 						 & ESO & 00-FEB-02 & 2.0 &  \\
B0430--278 & 1.630 &  21.3  &  CSS    &  $<17.0$ &0.3  &   0.76  & VLT & 99-NOV-11,DEC-12 & 2.4 &S \\
B0447--230 & 2.140 &  17.5  &  CSS    &     5.1  &1.48 &   0.83  & VLT & 99-NOV-13,15& 2.4 & S\\
B0451--282 & 2.560 &  17.8  &  $>5$   &     U~~  &0.93 &   0.06  & AAT & 97-FEB-09 & 2.4 & S\\
           & & & & & & 						 & AAT & 99-APR-12 & 2.4 & \\
B0522--215 & 1.820 &  22    &  CSS    &    15.3  &1.66 &   1.00  & \dots &\\
B0549--213 & 2.245 &  19.1  &  0.55   &    28.7  &0.00 &   0.83  & AAT & 99-APR-11 & 2.4 & S\\
           & & & & & & 						 & AAT & 97-FEB-09 & 2.4 &  \\
B1019--227 & 1.550 &  21.1  &  CSS    &    18.7  &0.67 &   0.98  & ESO & 00-FEB-02 & 2.0 & S\\
            & & & & & & 					 & AAT & 99-APR-11 & 2.4 & \\
B1025--264 & 2.665 &  17.5  &  5.61   &    77.3  &$-0.7$ & 0.63  & AAT & 97-FEB-09 & 2.4 & S\\
B1043--291 & 2.128 &  18.6  &  $>5$   &     8.0  &1.52 &   0.19  & AAT & 99-APR-11 & 2.4 & S\\
            & & & & & & 					 & AAT & 97-FEB-09 & 2.4 & \\
B1106--227 & 1.875 &  20.8  &  CSS    &    10.9  &1.54 &   0.73  & AAT & 99-APR-11,12 & 2.4 & S  \\
             & & & & & & 					 & AAT & 96-MAR-21,22 & 1.2 & \\
B1114--220 & 2.282 &  20.2  &  CSS    &     7.9  &2.77 &   0.72  & AAT & 96-MAR-21,22 & 1.2 & S \\
             & & & & & & & 					   AAT & 97-FEB-09    & 2.4 &   \\
             & & & & & & 					 & AAT & 99-APR-11,12 & 2.4 & \\
             & & & & & & 					 & VLT & 99-DEC-11 & 2.4 &  \\
B1212--275 & 1.656 &  19.6  &   $?$   &    29.6  &0.72 &   0.96  & ESO & 00-FEB-01 & 2.0 & S \\
             & & & & & & 					 & AAT & 97-FEB-09 & 2.4 &  \\
             & & & & & & 					 & AAT & 99-APR-12 & 2.4 &  \\
B1256--243 & 2.263 &  17.6  &  $>3$   &    65.2  &1.41 &   0.37  & ESO & 00-FEB-01 & 2.0 & S\\
             & & & & & & 					 & AAT & 99-APR-11,12 & 2.4 & \\
B1311--270 & 2.186 &  19.3  &  0.24   &   156.4  &1.15 &   0.82  & AAT & 99-APR-11 & 2.4 & S  \\
B1355--215 & 1.604 &  19.9  &  0.84   &    35.6  &1.52 &   0.80  & AAT & 99-APR-11,12& 2.4& S \\
B2122--238 & 1.756\tablenotemark{\ast} 
		   &  17.8  &  0.50   &    13.4  &0.76 &   0.71  & \dots &\\
B2128--208 & 1.610 &  20.0  &  CSS    &     6.9  &\dots &  0.97  & \dots &\\
B2158--206 & 2.249\tablenotemark{\ast} 
		   &  20.1  &  CSS    &     2.9  &0.63 &   0.87  & VLT & 99-OCT-17,NOV-12 & 2.4 &S \\
            & & & & & & 					 & AAT & 95-OCT-20 &  1.2 & \lya \\
B2210--257 & 1.831 &  17.9  &  $>3$   &     U~~  &1.10 &   0.18  & \dots &\\
B2211--251 & 2.500\tablenotemark{\ast}  
		   &  19.6  &  CSS    &    18.5  &1.59 &   0.98  & VLT & 99-OCT-17,NOV-12 & 2.4 & S \\
B2256--217 & 1.771\tablenotemark{\ast} 
                   &  19.9  &  0.085  &   277.8  &1.31 &   1.2   & \dots & \\
\tableline
\end{tabular}
\medskip \footnotesize

Columns--- (1) name; 
(2) mean emission-line redshift measured from low-resolution spectra; 
(3) COSMOS UKST magnitude ($b_{\rm J}$); 
(4) radio core-to-lobe luminosity ratio (measured at 10\,GHz in the restframe), 
or classification as CSS or GPS;
(5) radio linear size ($D$) in kpc (U=unresolved, beamed source); 
(6) optical spectral index ($\alpha_{\rm opt}$) as measured between 3500--10000\AA;
(7) radio spectral index, $\alpha_{0.408}^{4.86}$  measured between 408MHz and 4.86GHz;
(8) telescope;
(9) UT date of observations;
(10) spectral resolution FWHM;
(11) notes: where multiple spectra were obtained,  `S' denotes
those spectra shown in Figure \ref{grdspectra}.
The optical and radio data are taken from Kapahi et al. 1998 and 
Baker et al. 1999, except $\alpha_{\rm opt}$ for 
0447--230, 0451--282, 0522--215 and 1019--227 which are unpublished
measurements from de Silva et al. (2001, in preparation). 

\tablenotetext{\ast}{ Redshift revised, since Baker et al. 1999}
\end{minipage}
\end{table*}

\renewcommand{\arraystretch}{1.0}
\newpage
\begin{table*}
\begin{minipage}{17cm}
\caption[]{
MQS quasars with $0.7<z<1.0$ targetted with HST}
\label{lowtab}
\begin{tabular}{@{}cclcrccccl}
\\
\tableline
\small
MRC Quasar & $\langle z_e\rangle$ & $b_{\rm J}$ & $R$ & $D$ (kpc) &
$\alpha_{\rm opt}$ & $\alpha_{0.408}^{4.86}$  & UT Date & Time (s) & Notes \\
(1) & (2) & (3) & (4) & (5) & (6) & (7) & (8) & (9) & (10) \\
\tableline
\\ 
B0030--220 & 0.806 & 18.9 & 0.150 & 32   & 1.54 & 0.93 & 2000-Jun-20 & 5037 \\ 
B0106--233 & 0.818 & 20.1 & 0.25  & 21   & 0.54 & 0.85 & 2001-Feb-03 & 4738 \\ 
B0123--226 & 0.717 & 19.9 & 2.10  & 41   & 0.58 & 0.58 & 1999-Oct-16 & 4738\\ 
B0135--247 & 0.835 & 18.9 & $>2$  &\dots & 0.31 & 0.37 & 2001-Feb-05 & 5037\\ 
B0327--241 & 0.895 & 19.4 & $>1$  &\dots & 1.07 & 0.18 & 2000-Sep-20 & 4798\\ 
B0413--210 & 0.807 & 18.4 & 0.71  & 41   & 1.60 & 0.73 & 1999-Sep-05 & 5053 \\ 
B0437--244 & 0.834 & 17.5 & 0.098 & 1050 & 0.63 & 0.93 & 1999-Aug-13 & 2206 \\  
B0450--221 & 0.898 & 17.8 & 0.086 & 120  & 1.40 & 1.11 & 1999-Aug-16 & 2202\\ 
B1202--262 & 0.786 & 19.8 & 1.78  & 123  & 1.32 & 0.53 & 1999-Jul-30 & 4772\\ 
B1208--277 & 0.828 & 18.8 & 0.088 & 359  & 1.92 & 0.83 & 1999-Jun-28 & 5071 & Narrow \lya, \civ\ \\ 
B1217--209 & 0.814 & 20.2 & 0.057 & 246  & 0.17 & 0.94 & 1999-Aug-15 & 4738\\ 
B1222--293 & 0.816 & 18.5 & 0.38  & 243  & 0.66 & 0.76 & 1999-Jul-30 & 5079 \\ 
B1224--262 & 0.768 & 19.8 & CSS   & 5    & 1.49 & 0.83 & 1999-Aug-07 & 4772 \\ 
B1349--265 & 0.934 & 18.4 & CSS   &  2   & 2.50 & 0.63 & 1999-Aug-20 & 5079\\
B1355--236 & 0.832 & 19.9 & 0.098 & 133  & 1.17 & 0.95 & 1999-Jun-05 & 2198\\ 
B1359--281 & 0.802 & 18.7 & CSS   &$<12$ & 0.81 & 0.52 & 1999-Sep-07 & 5071\\ 
B2136--251 & 0.940 & 18.1 & GPS   &$<0.2$& 1.97 & 0.57 & 1999-May-30 & 5057\\ 
B2156--245 & 0.862 & 20.2 & CSS   & 1.9  & 2.05 & 0.87 & 1999-Sep-11 & 4738 & not detected \\
B2255--282 & 0.927 & 16.6 & $>5$  & \dots& 0.69 & $-0.23$ & 1999-May-22 & 2223 \\ 
 
\tableline
\end{tabular}
\medskip 

Columns--- (1) to (7) as for Table \ref{hightab};
$\alpha_{\rm opt}$ for 
B0437--244 was taken from de Silva et al. (2001, in preparation);
column (8) UT date of HST observation;
(9) total exposure time; (10) Notes.
 \end{minipage}
\end{table*}

\newpage

\begin{table*}
\begin{minipage}{15cm}
\caption[]{\civ\ emission and absorption measurements for $1.5<z<3.0$ MQS quasars}
\label{datatab}
\begin{tabular}{@{}ccccccrrrrll}
\\
\tableline
    & \multicolumn{3}{c}{Emission} && \multicolumn{6}{c}{Absorption} & \\
\cline{2-4} \cline{6-11}
MRC & \multicolumn{2}{c}{CIV\,1550}  & MgII\,2800 && \multicolumn{2}{c}{CIV\,1548}
&  CIV             &   MgII  & \multicolumn{2}{c}{CIV (\AA)} & \\
quasar    & $\lambda_{e}$ (\AA) & $z_e$       & $z_e$ && $\lambda_{a}$ (\AA) & 
$z_a$       & \multicolumn{2}{c}{$\Delta v$ (\kms)}  & $W_{\rm abs}$  & $\Delta W_{\rm abs}$ & 
\\ 
(1) & (2) & (3)  & (4) && (5) & (6) & (7) & (8)  & (9)  & (10) & 
\\ 
\tableline
 \\
B0133--266 & 3919 & 1.528 & 1.531 && 3955 & 1.555 &$-3200$&$-2800$ & 0.3 & 0.1 & \\
B0136--231 & 4494 & 1.899 & 1.897 &&(4492)& 1.899 & 0     & $-200$ & 2.4 & 0.3 & \\
B0222--224 & 4030 & 1.600 & 1.599 &&(4032)& 1.604 &$-450$ &$-600$  & 4.6 & 0.4 & \\
B0328--272 & 4343 & 1.802 & 1.818\rlap{:} && 4360 & 1.816 &$-1500$& 200\rlap{:} 	  & 2.1 & 0.1 & \\
B0413--296 & 4049 & 1.612 & 1.616 && 4041 & 1.610 & 250   & 700 	  & 1.5 & 0.1 & \\
B0430--278 & 4081 & 1.633 &  \dots&& 4075 & 1.632 & 100   & \dots  & 1.4 & 0.2 & \\
B0447--230 & 4870 & 2.142 & \dots && 4866 & 2.142 & 0     & \dots  & 5.2 & 0.2 & \\
B0451--282 & 5516 & 2.559 & \dots && \dots & \dots &\dots  & \dots  &$<0.2$& \dots& \\
           &      &       &       && 5419 & 2.500 &5000\tablenotemark{\ast}
							 & \dots  & 0.9 & 0.2 &  \\
B0549--213 & 5027 & 2.243 & \dots && 5000 & 2.229 &1300   & \dots  & 0.9 & 0.2 & \\
B1019--227 & 3948 & 1.547 & 1.544\rlap{:} && 3934 & 1.541 &700    & 350\rlap{:}	  & 1.4 & 0.1 & \\
B1025--264 & 5679 & 2.664 & 2.592 && \dots & \dots &\dots  & \dots  &$<0.2$& \dots& \\
           &      &       &       && 5561 & \dots &5900\tablenotemark{\ast}   
							 & \dots  & 0.6 & 0.2 & \\
B1043--291 & 4844 & 2.125 & 2.125 && \dots & \dots &\dots  & \dots  &$<0.2$& \dots& \\
           &      &       &       && 4838 & 2.069 &5400\tablenotemark{\ast}   
							 & 5400   & 1.3 & 0.3 & \\
B1106--227 & 4460 & 1.877 & 1.875 &&(4429)& 1.859 & 1900  & 1650	  & 1.3 & 0.3 & \\
B1114--220 & 5092 & 2.285 & 2.288 && 5104 & 2.297 &$-1100$&$-800$  & 3.2 & 0.1 & \\
B1212--275 & 4114 & 1.654 & 1.645 &&\dots & \dots &\dots  & \dots  &$<0.2$& \dots& \\
B1256--243 & 5061 & 2.265 & 2.262 && 4922 & 2.179 &7900\tablenotemark{\ast}   
							 & 7600   &$<0.2$ & \dots & \\ 
B1311--270 & 4949 & 2.194 &\dots &&\dots & \dots &\dots  & \dots  &$<0.1$& \dots& \\ 
B1355--215 & 4043 & 1.608 & 1.610 && 4039 & 1.609 &$-100$ & 100	  & 3.6 & 0.3 & \\
B2158--206 & 5022 & 2.240 & 2.256 && 4981 & 2.217 & 2150  & 3600	  & 1.1 & 0.1 & \\
B2211--251 & 5412 & 2.492 & \dots &&(5405)& 2.487 & 450   & \dots  & 1.8 & 0.2 & \\
\tableline
\end{tabular}
\smallskip

\medskip 
Columns---
(1) MQS quasar; (2) Peak wavelength ($\lambda_{e}$) and (3) redshift ($z_e$) for \civ\gl1550 
emission line from these observations; (4) redshift for \mgii\gl2800 emission line 
from low-resolution spectra in Baker et al. (1999) (or de Silva et al. in prep.,
with colons); (5) Absorption wavelength and (6) redshift ($\lambda_{a}$, $z_a$) for
\civ\gl1548 line (or central wavelength of blended doublet given in parentheses);
(7) \& (8) velocity offset ($\Delta v$ in \kms) of \civ\gl1548 absorption line 
relative to the \civ\gl1550 and \mgii\gl2800 emission lines, respectively
(positive means absorption is blueshifted relative to emission);
(9) restframe equivalent width  $W_{\lambda}$ of \civ\ doublet in absorption
(or limits) and (10) measurement uncertainties ($\Delta W$). 

\smallskip
\tablenotetext{\ast}{ $\Delta v \ge 5000$ \kms; 
	velocity exceeds definition of associated absorption 
(see Section \ref{sec:relvel}). An upper limit is also given for non-detection of 
associated absorption within 5000\,\kms.
}
\end{minipage}
\end{table*}

\newpage
\begin{table*}
\begin{minipage}{17.5cm}
\caption[]{\civ\ emission and absorption measurements for 
$0.7<z<1.0$ MQS quasars}
\label{hstdatatab}
\begin{tabular}{@{}ccccccccrrrrllll}
\\
\tableline
\footnotesize
    & \multicolumn{4}{c}{Emission} && \multicolumn{7}{c}{Absorption} & \\
\cline{2-5} \cline{7-13} 
MRC & \multicolumn{2}{c}{CIV\,1550}  & MgII &  [OII] &&
\multicolumn{2}{c}{CIV\,1548}
&  CIV             &   MgII  & [OII] && \multicolumn{2}{c}{CIV (\AA)} & \\
quasar    & $\lambda_{e}$(\AA) & $z_e$  & $z_e$     & $z_e$ && 
$\lambda_{a}$(\AA) & 
$z_a$       & \multicolumn{3}{c}{$\Delta v$ (\kms)}  & $W_{\rm abs}$  & $\Delta W_{\rm abs}$ & 
\\ 
(1) & (2) & (3)  & (4) & (5) && (6) & (7) & (8)  & (9)  & (10) & (11) & (12) & 
\\ 
\tableline
\\
B0030--220 & 2801 & 0.807 &\dots  & 0.804 && 2804 & 0.811 &$-700$ &\dots  &$-1200$& 1.7 & 0.6 &N \\
B0106--233 & 2821 & 0.820 & \dots & 0.817 &&(2820)&(0.819)& 100   &\dots  &$-300$ & 1.0 & 0.3 & \\
B0123--226 & 2667 & 0.721 & 0.715 & 0.720 && \dots& \dots &\dots  &\dots  &\dots  & $<0.3$&\dots & \\
B0135--247 & 2842 & 0.834 & 0.837 & 0.834 && 2842 & 0.836 &$-400$ & 200   &$-400$ & 0.7 & 0.1 &N \\ 
B0327--241 & 2931 & 0.891 & 0.898 & \dots && 2930 & 0.893 &$-200$ & 900   &\dots  & 0.2 & 0.1 & \\ 
B0413--210 & 2802 & 0.808 & 0.809 & 0.807 && \dots& \dots &\dots  &\dots  &\dots  & $<0.3$&\dots & \\ 
B0437--244 & 2839 & 0.832 & 0.825 & \dots && 2832 & 0.829 & 400   &$-700$ & \dots & 0.3 & 0.15 & \\  
B0450--221 & 2944 & 0.899 & 0.906 & 0.900 && 2940 & 0.899 &   0   & 1100  & 200   & 1.6 & 0.3 & \\
B1202--262 & 2772 & 0.788 & 0.789 & \dots && \dots& \dots &\dots  &\dots  &\dots  & $<0.2$&\dots & N \\
B1208--277 & 2834 & 0.828 & 0.825 & 0.829 && \dots& \dots &\dots  &\dots  &\dots  & \dots &\dots & N \\
B1217--209 & 2821 & 0.820 & 0.816 & \dots && \dots& \dots &\dots  &\dots  &\dots  & $<0.8$ &\dots & \\  
B1222--293 & 2816 & 0.817 & 0.817 & 0.820 &&\dots & \dots &\dots  &\dots  &\dots  & $<0.2$&\dots & \\
B1224--262 & 2751 & 0.775 & 0.773 & 0.772 && 2739 & 0.769 & 1000  & 700   & 500    & 3.9 & 0.8 & N \\
B1349--265 & 2992 & 0.930 & 0.923 & 0.925 && 2979 & 0.924 & 1000  &$-200$ & 100   & 3.8 & 0.6 & N \\ 
B1355--236 & 2841 & 0.833 & 0.840 & 0.834 && \dots& \dots &\dots  &\dots  &\dots  & $<0.8$&\dots & \\
B1359--281 & 2796 & 0.804 & 0.804 & 0.802 && 2797 & 0.807 &$-500$ &$-500$ &$-800$ & 2.8 & 0.6& \\ 
B2136--251 & 3005 & 0.939 & 0.946 & 0.939 &&\dots & \dots &\dots  &\dots  &\dots  & $<0.8$&\dots &N \\ 
B2255--282 & 2987 & 0.927 & 0.919 & 0.925 &&\dots & \dots &\dots  &\dots  &\dots  & $<0.3$&\dots & \\
 \\ 
\tableline
\end{tabular}
\footnotesize
Columns--- as Table \ref{datatab}, with addition of $z_e$ \oii\ (Col. 5) 
and $\Delta v$ (Col. 10)
relative to the redshift defined by the narrow \oii\gl3727 
emission line (visible in optical spectra, 
Baker et al. 1999, de Silva et al., in prep.). N=notes in text. 
\end{minipage}
\end{table*}

\normalsize
\renewcommand{\arraystretch}{1.0}
\begin{table*}
\begin{minipage}{16cm}
\caption[]{Statistics of strong absorbers in three quasar types}
\label{restab}
\begin{tabular}{@{}cccccccccccrrclll}
\\
\tableline
& \multicolumn{3}{c}{CSS} & & \multicolumn{3}{c}{LSS} & &
\multicolumn{3}{c}{CFS} \\
\cline{2-4} \cline{6-8} \cline{10-12} 
& low-$z$ & high-$z$ & Comb & 
& low-$z$ & high-$z$ & Comb &
& low-$z$ & high-$z$ & Comb \\ 
\tableline
\\
Median \ewa(\AA)& 3.3 &1.6 &2.8 &&$<0.8$&0.9 &$<0.8$&&0.2&$<0.2$&$<0.3$   \\
$N($total)      & 4   & 8  & 12 && 10     & 9   &19 && 3  & 3 &6\\
$N($\ewa$>1$\AA)& 3   & 8  & 11 && 4      & 4   &8  && 0  & 0 &0\\
\% (\ewa$>1$\AA)& 75 & 100  & 92 && 40     & 44  &42 && 0  & 0 &0\\
\tableline
\end{tabular}

Columns--- the quasars have been divided into three types: 
compact steep-spectrum
(CSS : $D<25$~kpc and \arad$>0.5$; as 
measured between 408\,MHz and 4.86\,GHz, and where $S_{\nu} \propto \nu^{-\alpha}$),
large steep-spectrum (LSS : $D>25$~kpc and \arad$>0.5$), and
compact flat-spectrum (CFS : \arad$<0.5$, all are compact
sources with $D<25$~kpc). Statistics are given for the
low-redshift ($0.7<z<1.0$), high-redshift ($1.5<z<3.0$) 
and combined samples. 

Rows--- the following for each type:
median equivalent width, \ewa; number of quasars with 
\ewa$>1$\AA\ and total number of quasars of that type, $N($total); 
and percentage of quasars with  \ewa$>1$\AA. 

\end{minipage}
\end{table*}

\end{document}